\documentclass[a4paper, 10pt]{article}
\usepackage{geometry}
\usepackage[utf8]{inputenc}
\usepackage{hyperref}
\usepackage{amsmath}
\usepackage{amssymb}
\usepackage{cleveref}
\usepackage{multirow}
\usepackage{float}
\usepackage{cleveref}
\usepackage{graphicx}
\usepackage{subcaption}
\usepackage{authblk}
\usepackage{color}

\DeclareMathOperator\arctanh{arctanh}
\newcommand{\kahler}{K\"ahler}

\title{\bf Gravitational Waves and Gravitino Mass in No-Scale Supergravity Inflation with Polonyi Term}
\author[1,2]{Miguel Crispim Rom\~ao\thanks{Email: \texttt{mcromao@lip.pt}}}
\author[1]{Stephen F. King\thanks{Email: \texttt{s.f.king@soton.ac.uk}}}

\affil[1]{School of Physics and Astronomy, University of Southampton. Southampton SO17 1BJ, United Kingdom}
\affil[2]{Laborat\'orio de Instrumenta\c{c}\~ao e F\'isica Experimental de Part\'iculas, Escola de Ciências, Departamento de F\'isica, Universidade do Minho, 
4701-057 Braga, Portugal}

\begin{document}
\maketitle

\begin{abstract}
    We study a No-Scale supergravity inflation model which has a non-minimal deformation of the \kahler\ potential and a Wess-Zumino superpotential extended by the inclusion of a Polonyi mass term. The non-minimal structure of the \kahler\ potential is responsible for an inflexion point that can lead to the production of gravitational waves at late stages of inflation, while the Polonyi term breaks supersymmetry at the end of inflation, generating a non-vanishing gravitino mass. After a thorough parameter space scan, we identify promising points for gravitational wave production. We then study the resulting gravitational wave energy density for this set of points, and we observe that the gravitational waves should be observable in the next generation of both space-based and ground-based interferometers. Finally, we show how the presence of the Polonyi term can be used to further boost the gravitational wave energy density, which is correlated with the gravitino mass. The code used for the scan and the numerical analysis is provided at \url{https://gitlab.com/miguel.romao/gw-and-m32-no-scale-inflation-polonyi}.
\end{abstract}

\section{Introduction}

With the first observation of Gravitational Waves (GW) by LIGO and Virgo collaborations~\cite{LIGOScientific:2016aoc,LIGOScientific:2016sjg,LIGOScientific:2017ycc}, we have witnessed a new era of Cosmology. Since then, multiple signals have been observed from astrophysical phenomena and, more recently, the NANOGrav collaboration has observed evidence for the existence of a stochastic GW background~\cite{NANOGRAV:2018hou,NANOGrav:2020bcs}. With multiple future GW experiments lined up, including space-based interferometers like LISA, BBO, and DECIGO~\cite{LISA:2017pwj,phinney2004big,Sato:2017dkf}, as well as next-generation ground-based interferometers such as the Cosmic Explorer and the Einstein Telescope~\cite{Reitze:2019iox,Maggiore:2019uih}, sensitivity to GW signals will soon include a multitude of phenomena beyond astrophysical events.

A particularly interesting source of GW that is not from astrophysical phenomena is the one associated with an enhancement of the scalar power spectrum (henceforth in this work refereed solely as power spectrum) during inflation. In this scenario, the first order scalar perturbations will drive second order perturbations that are themselves responsible for the production of GW~\cite{Acquaviva:2002ud,Ananda:2006af,Baumann:2007zm,Espinosa:2018eve}. In order to achieve this, the effective potential of the inflaton needs to exhibit a non-trivial feature where the inflaton considerably slows down, driving an enhancement of the power spectrum. Such a feature could be an inflexion point at a later stage of inflationary dynamics, where the inflaton slows down while traversing it due to the Hubble attrition. Such a scenario has been extensively studied recently within the context of Primordial Black-Holes (PBH) production~\cite{Nanopoulos:2020nnh,Stamou:2021qdk}, and more recently as a mechanism to generate GW in a No-Scale Supergravity~\cite{Spanos:2022euu}, which we will follow closely throughout this work.

Inflation~\cite{Guth:1980zm,Linde:1981mu,Mukhanov:1981xt,Albrecht:1982wi,Linde:1983gd} is a highly motivated framework that provides a solution for some of the problems unexplained by the Cosmological Standard Model~\cite{Linde:2007fr}, such as the flatness and horizon problems, the absence of cosmological relics, and the origin of cosmological fluctuations. Inflation is characterised by a period of sustained accelerated expansion due to a negative pressure. This can be obtained if all physical quantities are slowly varying. In particular, it can be described by the so-called slow-roll inflationary~\cite{Linde:1990flp,Lyth:1998xn} paradigm, where one (or many) scalar field(s) slowly evolve along a flat direction of the effective scalar potential. Currently, there is no model independent way of testing inflation, but observational data from the Planck satellite~\cite{Planck:2018jri} can constrain inflationary dynamics and rule out many candidate models of inflation. Many models of inflation have been proposed over the past years~\cite{Ringeval:2007am}, but most have been ruled out by the Cosmic Microwave Background (CMB) data from the Planck satellite. The remaining viable models fall inside a few classes, such as non-minimal gravity models, such as the $R^2$ Starobinsky inflation~\cite{Starobinsky:1980te,Starobinsky:1983zz}, Higgs and related inflation~\cite{Bezrukov:2009db,Linde:2011nh,Ferrara:2010in}, and hybrid low scale inflation~\cite{Copeland:1994vg,Dvali:1994ms}. While these classes of models survive Planck data, they are very sensitive to their possible Ultra-Violet (UV) completions, where generic corrections to the potential can lead to large corrections to the slow-roll parameters, spoiling inflation in what is known as the $\eta$-problem.

In the context of inflation, supersymmetry (SUSY) presents itself as a promising framework for models since in such models flat directions of the scalar potential are common, with UV corrections being naturally controlled by SUSY. For these reasons, SUSY has been a highly motivated ingredient for inflation~\cite{Ellis:1982ed,Ellis:1982dg,Ellis:1982ws}. Given that inflationary dynamics occurs so early in our Universe's history, and given the expected role that UV contributions might have, it is natural to study SUSY in its local form of Supergravity (SUGRA). In SUGRA, the F-term scalar potential receives contributions from two functions: the superpotential and the \kahler\ potential. It has been known that quadratic terms in the \kahler\ potential, which appear naturally if the \kahler\ potential is given by a minimal expansion in the (super)fields, will lead to the $\eta$-problem in SUGRA inflationary models. A solution to this problem is to consider the so-called No-Scale SUGRA, where the \kahler\ potential is given by a logarithmic form of the (super)fields, which are common in string theory compactifications, an idea that has gathered considerable interest lately~\cite{Ellis:2020lnc}.

In~\cite{Ellis2013-hj,Ellis:2013nxa}, Ellis, Nanopoulos, and Olive (ENO) have motivated a No-Scale \kahler\ potential arising from a non-compact $SU(2,1)/U(2)\times(1)$ coset compactification parametrised by two fields, a modulus $T$ and a chiral superfield $\Phi$. Using this \kahler\ potential in a Wess-Zumino framework for the superpotential, they showed that (for a given point in the parameter space) this SUGRA model is conformally equivalent to an $R+R^2$ Starobinsky inflation. Shortly after, we (CRK) have proposed a minimal extension of the Wess-Zumino superpotential by considering the existence of a Polonyi mass term~\cite{Romao2017-pj}. In our work, we showed that the Polonyi mass term leads to SUSY breaking, and that the viable region of the parameter space for inflation also set a bound on the gravitino mass, in the range of current and near future colliders. The phenomenology of SUSY breaking from our model has also been explored~\cite{King:2019omb} (for a general discussion of the status of no-scale SUGRA phenomenology see~\cite{Forster:2021vyz}). Recently, Nanopoulos, Spanos, and Stamou (NSS)~\cite{Nanopoulos:2020nnh} suggested that a deformation of the compact space considered by ENO would lead to an enhancement of the power spectrum that could be exploited to produce PBH. This new \kahler\ potential has been further been understood in more generic terms as breaking the non-compact coset symmetry of the compactification~\cite{Stamou:2021qdk}, and it was also shown that it can lead to the production of GW~\cite{Spanos:2022euu}.

In this work we revisit the No-Scale CRK model~\cite{Romao2017-pj}, which extended the one discussed by ENO by considering the presence of a Polonyi term in the superpotential, and we consider the impact of the more general \kahler\ potential with the deformation proposed by NSS. Our goal is to assess whether the phenomenology of the gravitino mass, a cornerstone feature of our previous model, is impacted by the presence of the NSS \kahler\ potential deformation, and, conversely, whether the phenomenology of the GW, a central feature of the NSS model, is impacted by the presence of the Polonyi term. In order to study this question, we perform a thorough analysis of the parameter space, constraining it by the Planck satellite data and viable inflation requirements, followed by a careful numerical analysis of the power spectrum and resulting GW energy density spectrum for a selection of promising points. We shall find that the gravitino mass in the CRK model is indeed correlated with the energy density of the GW signal.

This work is organised as follows. In~\cref{sec:noscale} we introduce the No-Scale Supergravity framework for inflationary models, where we set the notation, review main results, and present our model and its respective inflaton scalar potential in~\cref{sec:V}. Next, in~\cref{sec:scan}, we perform a thorough scan of the parameter space in order to identify regions in the parameter space that are in agreement with the Planck satellite constraints whilst producing viable inflation, and that show promising features that can lead to an enhancement of the power spectrum, which is needed for GW production. After identifying these promising points, in~\cref{sec:PR_and_GW} we start by computing the full numerical solution for the power spectrum in~\cref{sec:PR}, which we finally use to derive the present day energy density of the GW spectrum in~\cref{sec:GW}. In~\cref{sec:conclsions} we draw our conclusions and discuss future directions of work.

\section{The No-Scale Supergravity Inflation Framework\label{sec:noscale}}

We review the No-Scale SUGRA inflation framework to set notation, present our model, and derive all relevant formulae for later analysis.

\subsection{$\mathcal{N}=1$ Supergravity Preamble}

The main quantity in $\mathcal{N}=1$ SUGRA~\cite{Martin1997-bs,Brignole1997-vd} is the dimensionless real \kahler\ function, $G$, given by
\begin{equation}
    G = K(\Phi, \Phi^*) + \log |W(\Phi)|^2
\end{equation}
where $\Phi$ represents all the superfields in the theory, $K$ is the \kahler\ potential, itself a real function of the superfields, and $W$ is the superpotential, which is a holomorphic function of the superfields.  We also notice that we are following the customary SUGRA notations where we use the same symbol to refer to both the superfield and its scalar component, and the reduce Planck mass, $m_{Pl}$, is set to unity, i.e. $m_{Pl}  = 1$, and therefore all dimensionful quantities in this work are considered in units of $m_{Pl}$ unless explicitly given in GeV during the discussion.

The scalar F-term potential of the whole theory, $V_F$, is given by
\begin{align}
    V_F & = e^G( K^{i j^*} G_{i} G_{j^*} - 3)           \\
        & = K_{i^*  j} F^{i^*} F^{ j} - 3 e^K |W|^2 \ ,
    \label{eq:VF}
\end{align}
where the derivatives are taken with respect to the superfields with $\partial_i = \partial/\partial \Phi^i$, $\partial_{i^*} = \partial/\partial \Phi^{* i}$, and $ \partial_{i^*, j}= \partial^2/(\partial \Phi^{* i}\partial \Phi^{j})$, while the expression is evaluated on the scalar field component of the superfield, and
\begin{align}
    K_{i^* j} & = \partial_{i^*, j} K                                                  \\
    F^i       & =  - e^{K/2} K^{i j^*} (\partial_{j^*} W^* + W^* \partial_{j^*} K) \ ,
\end{align}
where $K_{i^* j}$ is the \kahler\ metric, $K^{i j^*}$ its inverse (i.e., $K^{i j^*}K_{j^* k} = \delta^i_j$), and $F^i$ is the SUGRA F-term. For gauge singlets, which will be focus of our work, the total scalar potential is given by $V=V_F$ as there is no D-term scalar potential contributions.

A very important observable in SUGRA is gravitino mass. In the case for vanishing cosmological constant, it can be written as
\begin{equation}
    m_{3/2}^2= e^K |W|^2 =\frac{1}{3} K_{i^* j} F^{i^*} F^{ j}\ ,
    \label{eq:m32}
\end{equation}
where the last equality is obtained from~\cref{eq:VF} by imposing $V=V_F=0$. Since is is non-vanishing only when $\langle F^i \rangle \neq 0$, it can be taken to the order parameter of SUSY breaking.

Next, we introduce the forms of the \kahler\ potential and superpotential that we will be studying for the rest of the paper, and how the shape of the potential is affected by different values of the its parameters.

\subsection{The No-Scale \kahler\ Potential}

The general form of the No-Scale \kahler\ potential that we will study is
\begin{equation}
    K = - 3 \log \left( T+T^* + f(\Phi,\Phi^*) \right) \ ,
    \label{eq:Kgeneral}
\end{equation}
where $T$ is a modulus field, to be stabilised by some other mechanism defined by the UV completion of the theory, and $\Phi$ corresponds to a hidden sector matter superfield, whose scalar part will drive inflation, while $f$ is a real function.

In the $(\Phi, T)$ basis,~\cref{eq:Kgeneral} leads to the \kahler\ metric
\begin{equation}
    K_{i^* j} = \frac{3}{(T+T^*+f)^2}  \begin{pmatrix}
        | \partial_\Phi f|^2-(T+T^*+f)\partial_{\Phi^*,\Phi}f & \partial_{\Phi^*}f \\
        \partial_{\Phi}f                                      & 1
    \end{pmatrix} \ ,
    \label{eq:KahlerMetric}
\end{equation}
and we note that an important entry is the $(1,1)$ element
\begin{equation}
    K_{\Phi^*\Phi}  = \frac{3}{\left(T+T^*+f\right)^2}\left(| \partial_\Phi f|^2-(T+T^*+f)\partial_{\Phi^*,\Phi}f \right) \ ,
    \label{eq:Kphiphi}
\end{equation}
which, once $T$ is stabilised, needs to accounted for when studying the canonically normalised field evolution.

Two forms of the \kahler\ potential have been thoroughly studied. In~\cite{Ellis2013-hj,Ellis:2013nxa} ENO, proposed the simple form for a space defined by the non-compact $SU(2,1)/SU(2)\times U(1)$ coset, parametrised by a modulus $T$ and a superfield $\Phi$ as
\begin{equation}
    K_{ENO} = - 3 \log \left(T+T^* - \frac{\Phi^*\Phi}{3}\right) \ ,
    \label{eq:KENO}
\end{equation}
and they showed that leads to Starobinsky inflation when paired with the a Wess-Zumino superpotential, which we will discuss in the next section. More recently, NSS~\cite{Nanopoulos:2020nnh} (and~\cite{Stamou:2021qdk}) extended the discussion to deformation of the space described above, given by a \kahler\ potential of the form
\begin{equation}
    K_{NSS} = - 3 \log \left(T+T^* - \frac{\Phi^*\Phi}{3} + a e^{-b(\Phi+\Phi^*)^2} (\Phi+\Phi^*)^4\right) \ ,
    \label{eq:KNSS}
\end{equation}
where $a$ and $b$ are free parameters, and it was shown that the new exponential factor inside the logarithm will produce a feature in the potential which can take the form of an inflexion point from which primordial black holes and gravitational waves can be produced~\cite{Nanopoulos:2020nnh,Stamou:2021qdk,Spanos:2022euu}. For both \kahler\ potentials, the modulus $T$ is assumed stabilised and real, for which we take
\begin{equation}
    \langle T \rangle = \langle T^* \rangle = \frac{c}{2} \ ,
    \label{eq:Tvev}
\end{equation}
with vanishing derivatives.

\subsubsection{Kinetic Term and Canonically Normalised Field}

An immediate consequence of the \kahler\ metric,~\cref*{eq:KahlerMetric}, not being diagonal is that the field $\Phi$ is not canonically normalised. Taking the modulus $T$ to be stabilised as~\cref{eq:Tvev}, the kinetic term of $\Phi$ has a factor given by the $(1,1)$ entry of the \kahler\ matrix,~\cref{eq:Kphiphi}. In order to proceed, we need to identify a canonically normalised field which will play the role of the inflaton. Consider the complex field
\begin{equation}
    \chi = \frac{1}{\sqrt{2}}(x+ i y)
    \label{chi}
\end{equation}
as the canonically normalised field associated with $\Phi$,
\begin{equation}
    \mathcal{L}_{Kin}  \supset K_{\Phi^* \Phi} \partial_\mu \Phi^* \partial^\mu \Phi = \partial_\mu \chi^* \partial^\mu \chi = \frac{1}{2} \partial_\mu x \partial^\mu x +\frac{1}{2} \partial_\mu y \partial^\mu y \ ,
\end{equation}
and assume now that the imaginary part has a vanishing vacuum expectation value  (VEV) and it is not dynamic, so that we can focus on
$\phi = \Re(\Phi)$. From the previous equation, we see that the field $x$, which we will identify as the inflaton, is now the canonically normalised version of $\phi$, with
\begin{equation}
    \partial_\mu x = \frac{\partial x}{ \partial \phi} \partial_\mu \phi \ ,
\end{equation}
where the map between the two, $x = x(\phi)$, is given by integrating the exchange of variables coefficient given by the \kahler\ metric entry
\begin{equation}
    \frac{\partial x}{\partial \phi} = \sqrt{2 K_{\Phi^* \Phi}\big|_{\Phi=\phi}} \ .
    \label{eq:dxdphi}
\end{equation}

For the case of the ENO \kahler\ potential,~\cref{eq:KENO}, it is possible to perform the integration~\cref{eq:dxdphi} and obtain a closed-form expression
\begin{equation}
    \phi_{ENO} = \sqrt{3 c} \tanh \left(\frac{x}{\sqrt{6}}\right) \ .
    \label{eq:phieno}
\end{equation}
where $c$ is the modulus VEV in \cref{eq:Tvev}
However, for more complicated \kahler\ potentials, such as the NSS~\cref{eq:KNSS},~\cref{eq:dxdphi} can only be integrated numerically.

\subsection{Superpotential}

The second ingredient required to define the scalar potential is the superpotential. In~\cite{Ellis2013-hj,Ellis:2013nxa}, ENO assumed the minimal Wess-Zumino superpotential
\begin{equation}
    \label{eq:Weno}
    W_{ENO} = \frac{\hat\mu}{2} \Phi^2 - \frac{\lambda}{3} \Phi^3 \ ,
\end{equation}
which, with~\cref{eq:KENO}, produces a scalar potential where the inflaton is identified with $x$, as in the previous section, producing a Starobinsky-like inflation. In fact, this superpotential and~\cref{eq:KENO} have an exact Starobinsky limit for $\hat \mu = \mu/\sqrt{c/3}$ and $\lambda = \mu/3$. This superpotential was also considered by NSS paired with their proposed \kahler\ potential~\cref{eq:KNSS}.

Later, we have proposed extending the Wess-Zumino superpotential with the addition of the a Polonyi mass term~\cite{Romao2017-pj}
\begin{equation}
    \label{eq:Wcrk}
    W_{CRK} =M^2 \Phi + W_{ENO}  \ .
\end{equation}
The addition of the extra Polonyi mass term shifts the VEV of the inflaton at the end of inflation to a non-vanishing configuration, $\langle x \rangle \neq 0$, producing a non-vanishing gravitino mass, which is bounded from above for the regions of the parameter space that produce good inflation. We will revisit this result in the next section.

\subsection{Scalar F-Term Potential\label{sec:V}}

Regardless of the choice of superpotential, the generic No-Scale \kahler\ potential from~\cref{eq:Kgeneral}, leads to a scalar F-term potential for $\phi$ of the form
\begin{equation}
    \label{eq:generalpotential}
    V = \frac{ 1}{-3 \partial_{\Phi^*,\Phi}f
    } \frac{1}{\left(f+T+T^*\right)^2} |\partial_\Phi W|^2 \ \Big|_{\Phi=\Phi^*=\phi},
\end{equation}
and we have organised the expression as the product of three terms, where the last two are always positive semi-definite, but the first can take negative values for a generic real function $f=f(\Phi,\Phi^*)$. Note, however, that for the ENO \kahler\ potential,~\cref{eq:KENO}, one has $f=-|\Phi|^2/3$, and therefore the potential is always positive semi-definite. The same cannot be said for the NSS \kahler\ potential,~\cref{eq:KNSS}.

We now turn to case of the NSS \kahler\ potential,~\cref{eq:KNSS}, with the CRK superpotential,~\cref{eq:Wcrk}, which will be the focus of our work. To simplify notation, we also use the following parameter redefinitions for reasons that will become clear shortly
\begin{align}
    \hat \mu & \equiv \mu \sqrt{c/3}                 \\
    M        & \equiv \sqrt{\mu d}  \label{eq:d_def} \\
    \lambda  & \equiv \mu l \ , \label{eq:l_def}
\end{align}
where the replacement of $ \hat \mu$, $M$ and $\lambda $ by three new parameters
$\mu$, $d$ and $l$ does not change the total parametric freedom, which also includes the  \kahler\  parameters
$a,b$ in~\cref{eq:KNSS} and the modulus VEV $c$. This corresponds to six real parameters in total, although $\mu^2$ can be factored out of the scalar potential which depends on the five real parameters $a,b,c,d,l$, as we now discuss.

With these, we can obtain the final scalar F-term potential for the non-canonically normalised inflaton field, $\phi = \Re(\Phi)$,
\begin{equation}
    V = \frac{1}{1-24 a \phi ^2 e^{-4 b \phi ^2} \left(4 b \phi ^2 \left(8 b \phi ^2-9\right)+6\right)} \frac{1}{\left(c-\frac{\phi ^2}{3}+16 a e^{-4 b \phi ^2} \phi ^4\right)^2} \mu^2( d +  \sqrt{c/3} \phi -  l \phi^2 )^2 \ ,
    \label{eq:VFphi}
\end{equation}
which can be shown to reproduce our previous results for $a=0$ (with the ENO results obtained by further considering $d=0$), as well as exhibiting a Starobinsky limit for $l=1/3$\footnote{The expression~\cref{eq:VFphi} differs to the one in~\cite{Nanopoulos:2020nnh}. After contacting the authors, we were able to identify the source of the mismatch from their side.}. Furthermore, we can factorise $\mu^2$ as a result of the parametric redefinitions listed above.

\begin{figure}[H]
    \centering
    \begin{subfigure}{0.32\textwidth}
        \includegraphics[width=\linewidth]{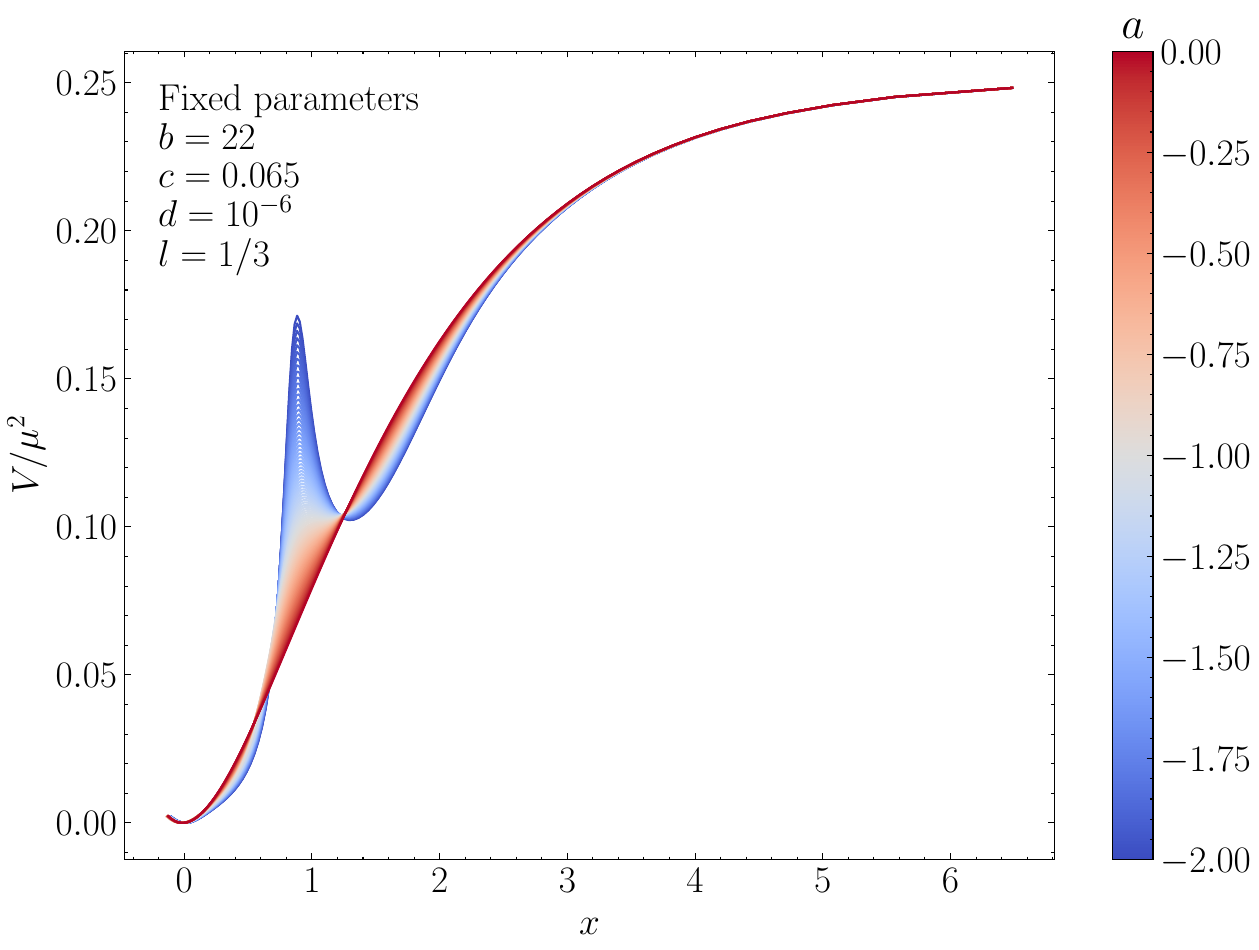}
        \caption{Varying $a\in [-2,0]$.}
        \label{fig:potential_shape_a}
    \end{subfigure}
    \begin{subfigure}{0.32\textwidth}
        \includegraphics[width=\linewidth]{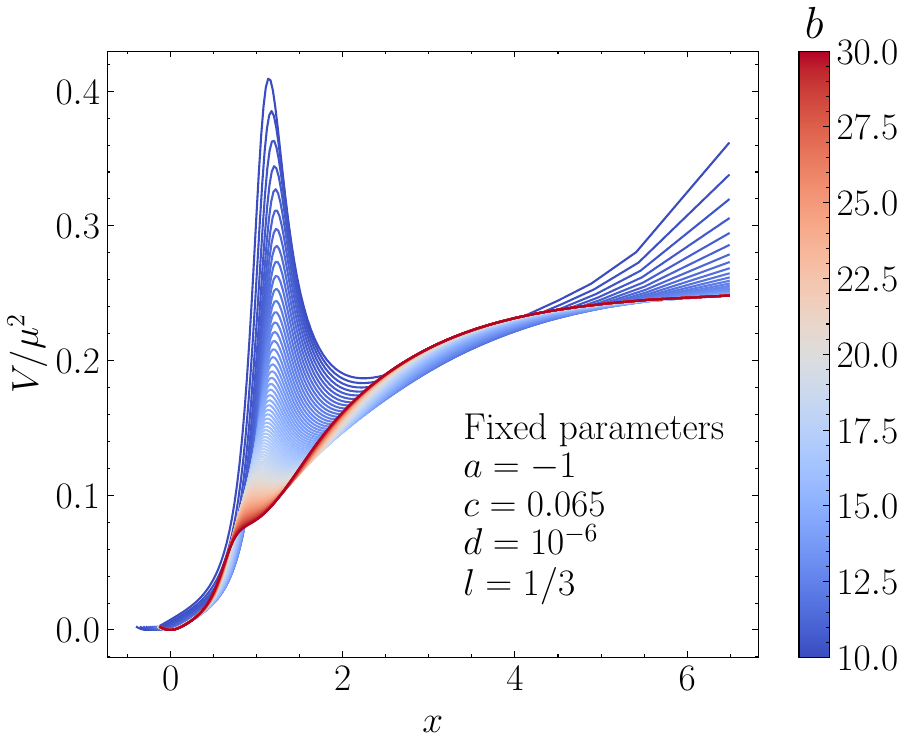}
        \caption{Varying $b\in [10,30]$.}
        \label{fig:potential_shape_b}
    \end{subfigure}
    \begin{subfigure}{0.32\textwidth}
        \includegraphics[width=\linewidth]{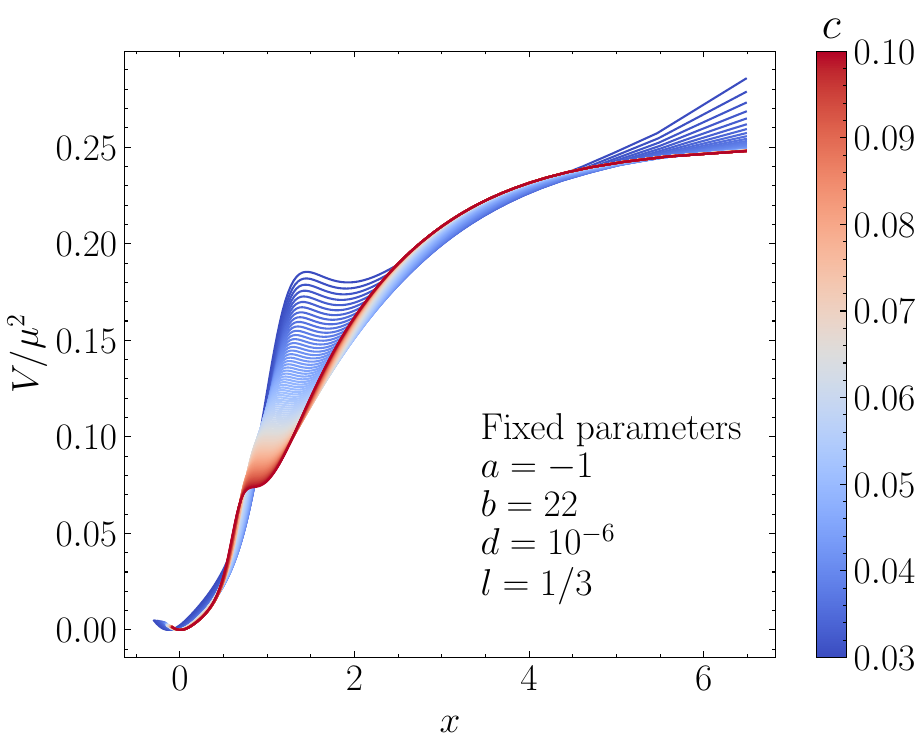}
        \caption{Varying $c\in[0.03,0.1]$.}
        \label{fig:potential_shape_c}
    \end{subfigure}
    \begin{subfigure}{0.32\textwidth}
        \includegraphics[width=\linewidth]{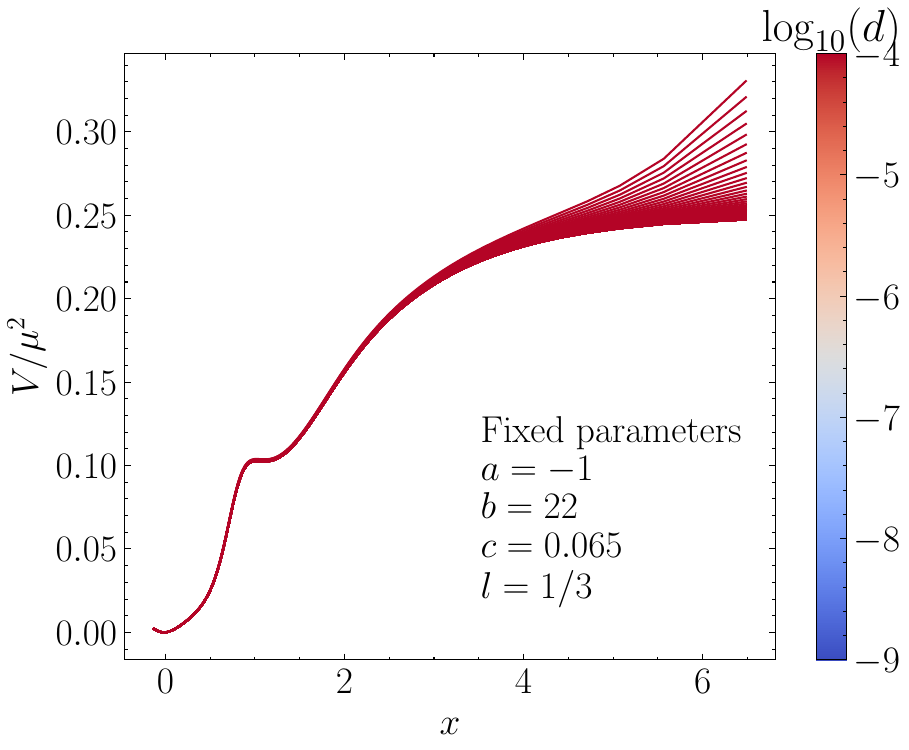}
        \caption{Varying $d\in [10^{-9}, 10^{-4}]$.}
        \label{fig:potential_shape_d}
    \end{subfigure}
    \begin{subfigure}{0.32\textwidth}
        \includegraphics[width=\linewidth]{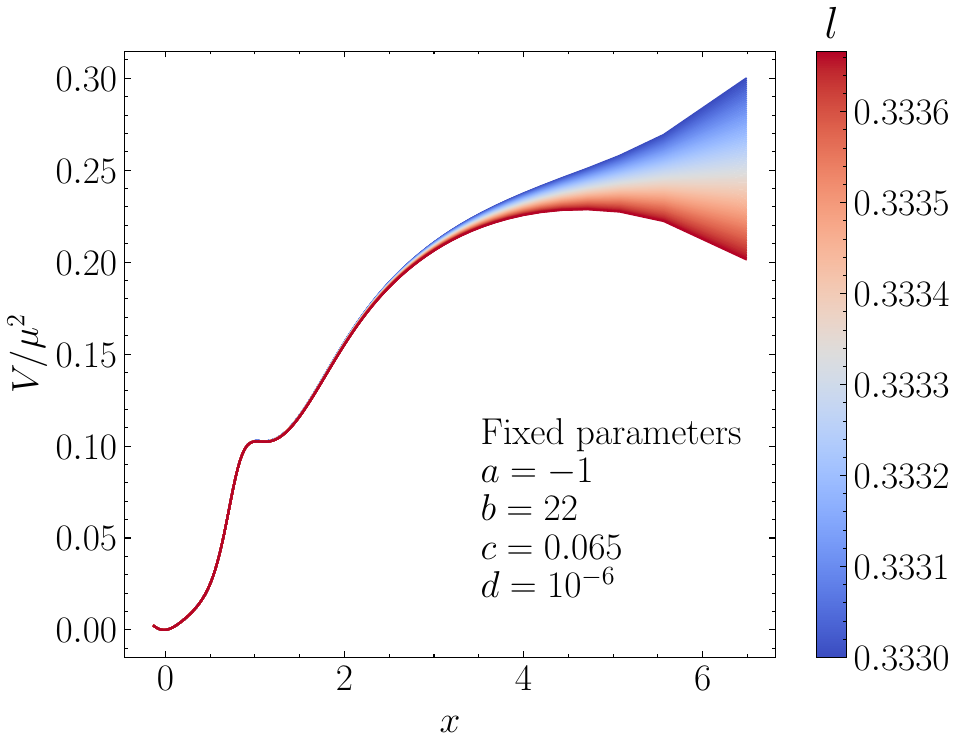}
        \caption{Varying $l \in \frac{1}{3} \times [0.999, 1.001]$.}
        \label{fig:potential_shape_l}
    \end{subfigure}
    \caption{The inflaton potential displays a kink structure which depends sensitivity on the different parameters. When fixed, the parameters take the values $a=-1$, $b=22$, $c=0.065$, $d=10^{-4}$, $l=1/3$.}
    \label{fig:potential_shape_parameters}
\end{figure}

With the expression for the scalar potential of the non-canonically normalised field $\phi$,~\cref{eq:VFphi}, and the change of variables from $\phi$ to the canonically normalised inflaton field, $x$, given by~\cref{eq:dxdphi}, we can plot the potential for different values of the parameters to assess their impact in the potential shape. In~\cref{fig:potential_shape_parameters} we present the potential where we take each parameter to vary in a certain range while keeping the remainder constant.

In the first row,~\cref{fig:potential_shape_a,fig:potential_shape_b,fig:potential_shape_c}, we observe how the shape of the potential varies as we vary the three parameters that arise from the \kahler\ potential $a$, $b$, and $c$, respectively. We observe that they all impact on the shape of the kink-like feature, although this always happens around $x\sim 1$. In~\cref{fig:potential_shape_a} we see that increasing the magnitude of $a$ will make a more pronounced feature in the potential, where the kink can become a wall. This is easily understandable, as the larger the magnitude of $a$ the bigger will be the contribution of the exponential part of $f$. Similarly, in~\cref{fig:potential_shape_b} we see that a smaller $b$ will lead to a more pronounced perturbation of the potential. This happens for the same reason as $a$, as $b$ tends to $0$ the exponential becomes leading inside logarithm in the \kahler\ potential. However, unlike $a$, the values of $b$ can impact the potential at around $x\sim 5$, where the potential plateaus and where inflation is expected to start, suggesting that too small values of $b$ will spoil the desired inflationary characteristics of the potential. With regards to the last \kahler\ potential parameter, $c$, in~\cref{fig:potential_shape_c} we see that for very small values we witness the kink becoming again more pronounced. This is understandable as for small $c$ the remaining terms of the \kahler\ potential become leading, increasing their impact in the shape and position of the kink. Like $b$, for extreme values of $c$ the potential is also considerably changed around $x\sim 5$, spoiling good inflationary dynamics. Finally, we notice that $b$ and $c$ appear to change the eight and the width of the kink, whereas $a$ can only turn the kink into a wall, but does not change its position.

In the second row,~\cref{fig:potential_shape_d,fig:potential_shape_l}, we observe how the shape of the potential is affected by the ranges of values for the superpotential parameters $d$, $l$, respectively. The first observation is that neither of these parameters affect the kink. This is unsurprising, as the kink is a feature arising from the new contributions to the \kahler\ potential proposed by NSS,~\cref{eq:KNSS}. In~\cref{fig:potential_shape_d} we see how the magnitude $d$, which is proportional to the Polonyi term c.f.~\cref{eq:d_def}, has limited impact on the shape of the potential unless it takes large values. This is in agreement with the findings in~\cite{Romao2017-pj}, where $M$ was found to have minimal impact on Starobinsky-like inflation unless the gravitino mass (which is mainly driven by $M$, as we will see below) is large. Finally, in~\cref{fig:potential_shape_l} we see the impact of $l$, c.f.~\cref{eq:l_def} in the shape of potential. Especially, we note how the plateau where inflation is expected to start is highly sensitive to a deviation of $l\neq 1/3$, with $l=1/3$ producing a Starobinsky-like inflationary model.

\subsubsection{Potential Positive Semi-Definiteness and Global Minimum}

As it can be seen from~\cref{eq:generalpotential}, the scalar potential for the inflaton is not always positive semi-definite. In particular, this means that we cannot identify the global minimum of the potential to be the field configuration where $V=0$ as it was done in~\cite{Romao2017-pj}. To understand this better, we notice that the potential is found to be positive semi-definite if the first factor in~\cref{eq:VFphi} is positive, i.e.
\begin{equation}
    1-24 a \phi ^2 e^{-4 b \phi ^2} \left(4 b \phi ^2 \left(8 b \phi ^2-9\right)+6\right) > 0 \ .
    \label{eq:positive_condition}
\end{equation}

\begin{figure}[H]
    \centering
    \includegraphics[width=0.9\textwidth]{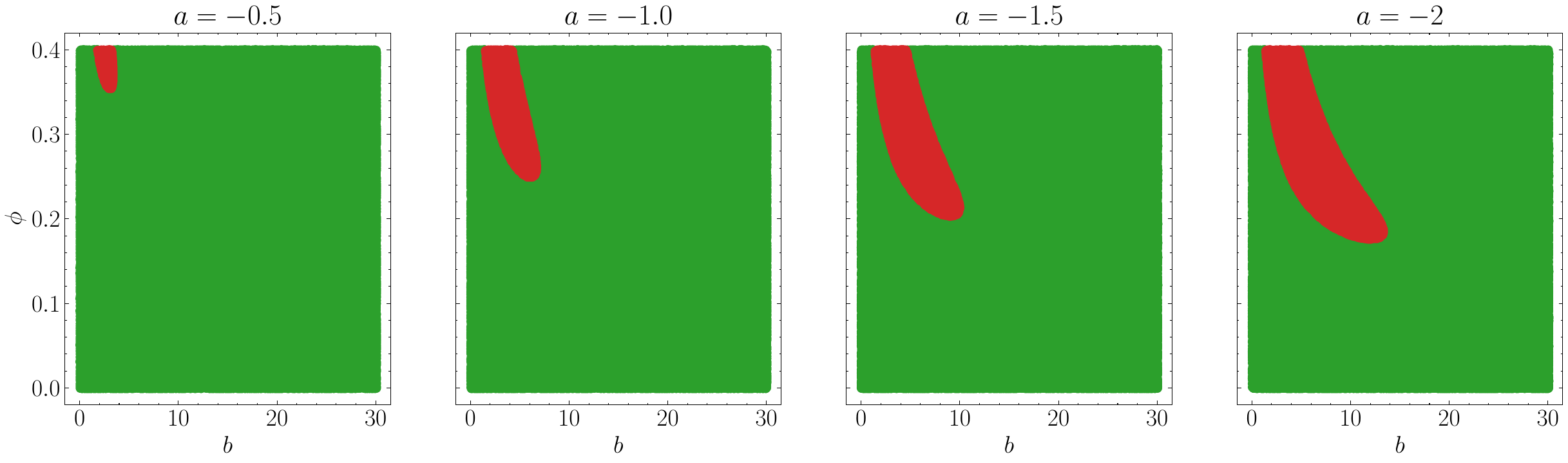}
    \caption{Positivity condition as a function of the parameter $b$ and the field $\phi$ value for different values of $a$. Red (green) indicates a negative (positive) value of the scalar potential.}
    \label{fig:positivity}
\end{figure}

In~\cref{fig:positivity} we study when~\cref{eq:positive_condition} holds for different values of $a$, $b$, and $\phi$. We find that as long as $a\gtrsim -1.5$ and $ b\gtrsim15$, that the potential is positive semi-definite for all values of $\phi$ during inflation. In such case, the minimum of the potential is given by $V=0$, or conversely, c.f.~\cref{eq:generalpotential}, by
\begin{align}
    \partial_\Phi W \Big|_{\Phi=\langle \Phi \rangle} = 0 \ ,
\end{align}
which happens to be a simple quadratic equation in $\Phi$, with solution
\begin{equation}
    \langle \Phi \rangle = \frac{\hat \mu \pm \sqrt{\hat \mu^2 + 4 M \lambda}}{2 \lambda} \ ,
\end{equation}
or, alternatively using the parameter redefinitions and focusing on $\phi$ we have
\begin{equation}
    \langle \phi \rangle = \frac{1}{2 l} \left(\sqrt{\frac{c}{3}} \pm \sqrt{\frac{c}{3} + 4 l d}\right) \ .
    \label{eq:phi_vev}
\end{equation}

Therefore, for $a\gtrsim -1.5$ and $ b\gtrsim15$ the inflaton, which is identified as the canonically normalised version of $\phi$, will achieve the global minimum at $V=0$ and be stabilised at a non-vanishing VEV, breaking SUSY and producing a non-vanishing gravitino mass
\begin{equation}
    \langle \phi \rangle \neq 0 \Rightarrow \langle F^\Phi \rangle \neq 0 \Rightarrow m_{3/2} \neq 0 .
\end{equation}

\section{Parameter Space Scan: Planck Data and Inflation\label{sec:scan}}

As it is observable in~\cref{fig:potential_shape_parameters}, the non-trivial contribution in $K_{NSS}$,~\cref{eq:KNSS}, can lead to a kink-like feature in the potential that can become wall-like, rather than an inflexion point or a plateau. As a consequence, the
canonically normalised inflaton field, $x$, which is the real part of $\chi$ in~\cref{chi},
can get stuck in the kink at later e-fold times instead of rolling down to the global minimum. Therefore, in order to study the parameter space of our model, we need to dynamically evolve the inflaton from the top of the potential to the global minimum. To do this, we need to solve its equation of motion, which in e-fold time, $N$, is given by~\cite{Spanos:2022euu}
\begin{align}
    \frac{d^2x}{d N^2} + \left(3 - \frac{1}{2}\left(\frac{d x}{d N}\right)^2\right) \frac{d x}{d N} + \left(3 - \frac{1}{2}\left(\frac{d x}{d N}\right)^2\right)\partial_x \ln V = 0 \ .
    \label{eq:EOM_x}
\end{align}

The closed-form for the potential appearing in the last term is given by~\cref{eq:VFphi} is only known in terms of $\phi$, not of $x$. However, given the field redefinition transformation,~\cref{eq:dxdphi}, we can use the expression of the potential in terms of $\phi$ as long as we take into account the field redefinition transformation in the derivative with regards to $x$, i.e. by replacing
\begin{align}
    \partial_x \to  \frac{1}{\sqrt{2 K_{\Phi^*\Phi}\Big|_{\Phi=\phi}}}\partial_\phi \ .
    \label{eq:partial_x_to_partial_phi}
\end{align}
With this transformation,~\cref{eq:EOM_x} has now an explicit dependence on $\phi$, therefore we need to evaluate it jointly with the e-fold time evolution of $\phi$, which is given by
\begin{align}
    \frac{d \phi}{d N} = \frac{1}{\sqrt{2 K_{\Phi^*\Phi}\Big|_{\Phi=\phi}}}\frac{d x}{d N} \ .
    \label{eq:dphidn_dxdn}
\end{align}

To integrate the equations above we need to set the initial conditions. We integrate from the beginning inflation, which we set to be $N_i=0$, until $N=70$ so that the integration covers the desired number of e-folds, $\Delta N$, which should take values $\Delta N = N_f - N_i =  \int_{a_i}^{a_e} d \ln a  \simeq 50-60$. The beginning of inflation happens at the high plateau of the potential. This is identified with the pivot scale, $k^*=0.05\text{ Mpc}^{-1}$, i.e. $N^*\simeq N_i = 0$, which is the scale at which the CMB crosses the decreasing comoving Hubble radius. Therefore, the initial conditions for $x^*=x(N^*)$ are set such that it is given asymptotically by
\begin{align}
    x^* = \sqrt{6} \arctanh \left(\frac{\phi^*}{\sqrt{3 c}}\right) \ ,
    \label{eq:x_ic}
\end{align}
which is justified by the fact that for large values of $\phi$ we have $K_{NSS} \to K_{ENO}$ and so we recover the field redefinition~\cref{eq:phieno}. In turn, this sets an upper bound on $\phi^*\lesssim \sqrt{3 c}$. As for the derivative initial condition, we set it to the so-called slow-roll attractor, with the choice of sign such that the inflaton rolls to decreasing values of $x$
\begin{align}
    \left(\frac{d x}{d N}\right) ^* \simeq - \left| \frac{1}{V}\partial_x V \right|_{x=x^*} \ .
    \label{eq:dx_ic}
\end{align}

With the above integration of the equations of motion we will be able to assess whether a point of the parameter space produces the desired inflation. In addition to this, at the pivot scale we will compare the predictions for the scalar to tensor ratio, $r$, and the scalar tilt, $n_s$, which were measured by the Planck satellite to be (at 3 sigma)~\cite{Planck:2018jri}
\begin{align}
    r   & < 0.055                  \\
    n_s & \in [0.9536, 0.9782] \ .
    \label{eq:planck_observables}
\end{align}
These can be computed for a choice of parameters at top of the potential using the potential slow-roll parameters, $\epsilon_V$ and $\eta_V$,
\begin{align}
    r   & \simeq 16 \epsilon_V                   \\
    n_s & \simeq 1 - 6 \epsilon_V + 2 \eta_V \ ,
\end{align}
with
\begin{align}
    \epsilon_V & = \frac{1}{2}\left(\frac{\partial_x V}{V}\right)^2 \\
    \eta_V     & = \frac{\partial_{x,x}V}{V} \ ,
\end{align}
where both quantities can be computed using the expressions in terms of $\phi$ when taking~\cref{eq:partial_x_to_partial_phi}, and are evaluated at $x=x^*$, i.e. at $\phi=\phi^*$.

Both $n_s$ and $r$ are independent of $\mu$, as it can be factored out from $V$ and it cancels in the expressions above. However, the value for $\mu$ can be fixed by another Planck observable, which is the amplitude of the power spectrum, $P_{R}$, measure at the pivot scale. At the pivot scale, we can use the slow-roll approximation for $P_{R}$, which reads
\begin{align}
    P_{R,sr} \simeq \frac{1}{8\pi^2} \frac{H^2}{\epsilon_H} \ ,
    \label{eq:PRsr}
\end{align}
where we add the label $s.r.$ to indicate it as a slow-roll approximation, $H$ is the Hubble radius, and $\epsilon_H$ the Hubble $\epsilon$ parameter, given by
\begin{align}
    H^2        & = \frac{V}{3 - \epsilon_H}                       \\
    \epsilon_H & = \frac{1}{2} \left(\frac{d x}{d N}\right)^2 \ ,
\end{align}
where $dx/dN$ is obtained from the numerical integration\footnote{Alternatively, at the pivot scale the full slow-roll approximation can be taken $\epsilon_H \simeq \epsilon_V$ and $\epsilon_V \ll 1$ s.t. $P_{R,s.r.} \simeq (1/8\pi^2) V^2/(3\epsilon_V)$.}. The power spectrum is measured by Planck to be $P_{R}\big|_{k=k^*}\simeq 2.1 \times 10^{-9}$.

The last observable that we will study is the gravitino mass at the end of inflation, i.e.
\begin{align}
    m^2_{3/2} = e^K |W|^2\Big|_{\Phi=\langle\phi\rangle} \ ,
\end{align}
where we choose the minus sign in~\cref{eq:phi_vev} as it corresponds to a global minimum to the left of the maximum of the potential, i.e. in the direction where $\phi$ is rolling to as set by~\cref{eq:dx_ic}.

\subsection{Random Scan}

For the range of the parameters, we follow the discussion in~\cref{sec:V} and in~\cite{Romao2017-pj,Spanos:2022euu} to focus the scan on a promising region of the parameter space:
\begin{align}
    a      & \in [-1.5, -0.5]                       \\
    b      & \in [15, 30]                           \\
    c      & \in 0.065 \times [0.95, 1.05]          \\
    d      & \in 10^{[-8, -2]}                      \\
    l      & \in [0.33327, 1/3]                     \\
    \phi^* & \in \sqrt{3 c} \times [0.95, 1.00] \ .
\end{align}
We randomly sample $10^6$ points from this parameter space. For each point, we collect the values of the observables listed above, the final value of $N_f$ at which the numerical integration converges, and the value of the inflaton field, $x$, at the end of inflation, $\langle x \rangle_{end}$. This scan was performed in Mathematica, with the system of differential equations~\cref{eq:EOM_x,eq:dphidn_dxdn} solved jointly using Mathematica's NDSolve.

\begin{figure}[H]
    \centering
    \includegraphics[width=1\textwidth]{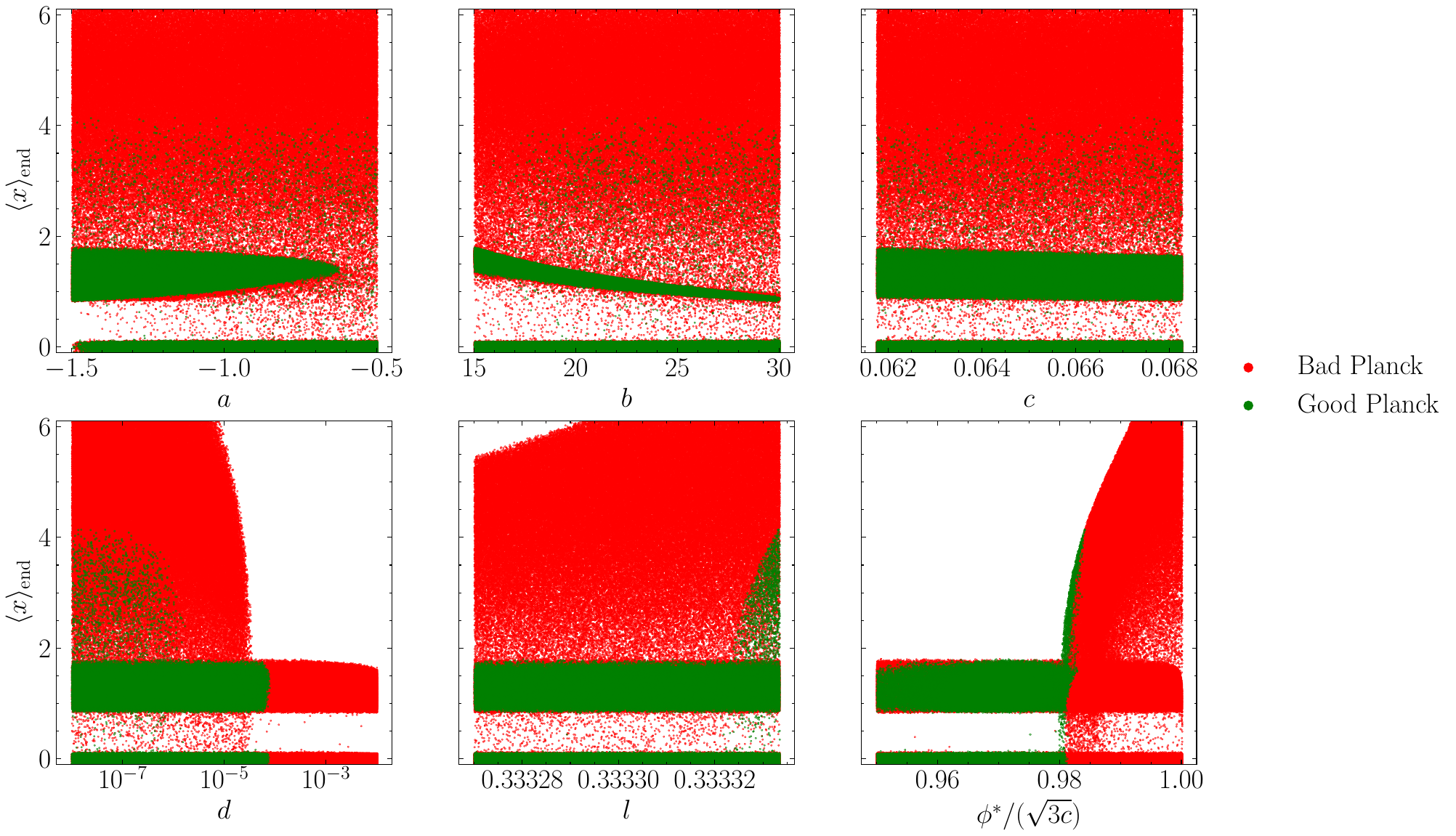}
    \caption{Inflaton field configuration, $\langle x\rangle_{end}$ at the end of the numerical integration versus the parameters of the scalar potential. Green (Red) points agree (disagree) with Planck constraints~\cref{eq:planck_observables}.}
    \label{fig:x_final_vs_parameters}
\end{figure}

The first noticeable feature observed in the scan, is that not all parameter points lead to the inflaton rolling down to the global minimum, as anticipated by the exploratory analysis in the previous section. In~\cref{fig:x_final_vs_parameters} we see how the final field configuration for $x$, $\langle x \rangle_{end}$, is affected by different values, and render each point either green or red depending on whether the predicted slow-roll observables agree with Planck or not, respectively. We see that there are two main clusters of points: those where the inflaton gets stuck at the kink, $ \langle x \rangle_{end} \gtrsim 1$, and those where the inflaton rolls down to the global minimum,  $ \langle x \rangle_{end} \lesssim 0.1$.
For the points that agree with Planck cosmological constraints, we can further see how the parameters $a$ and $b$ impact the value of $\langle x \rangle_{end}$. More precisely, we see how for $a\to0$ or $b\to\infty$ it is easier for the inflaton to roll down to the global minimum, as in these limiting cases the kink disappears completely. Conversely, for $a\to - \infty $ or $b\to 0$, the impact on the scalar potential from the non-trivial contribution in~\cref{eq:KNSS} becomes more pronounced and the kink goes from a plateau to a wall, preventing $x$ from rolling down to the global minimum.

\begin{figure}[H]
    \centering
    \includegraphics[width=1.0\textwidth]{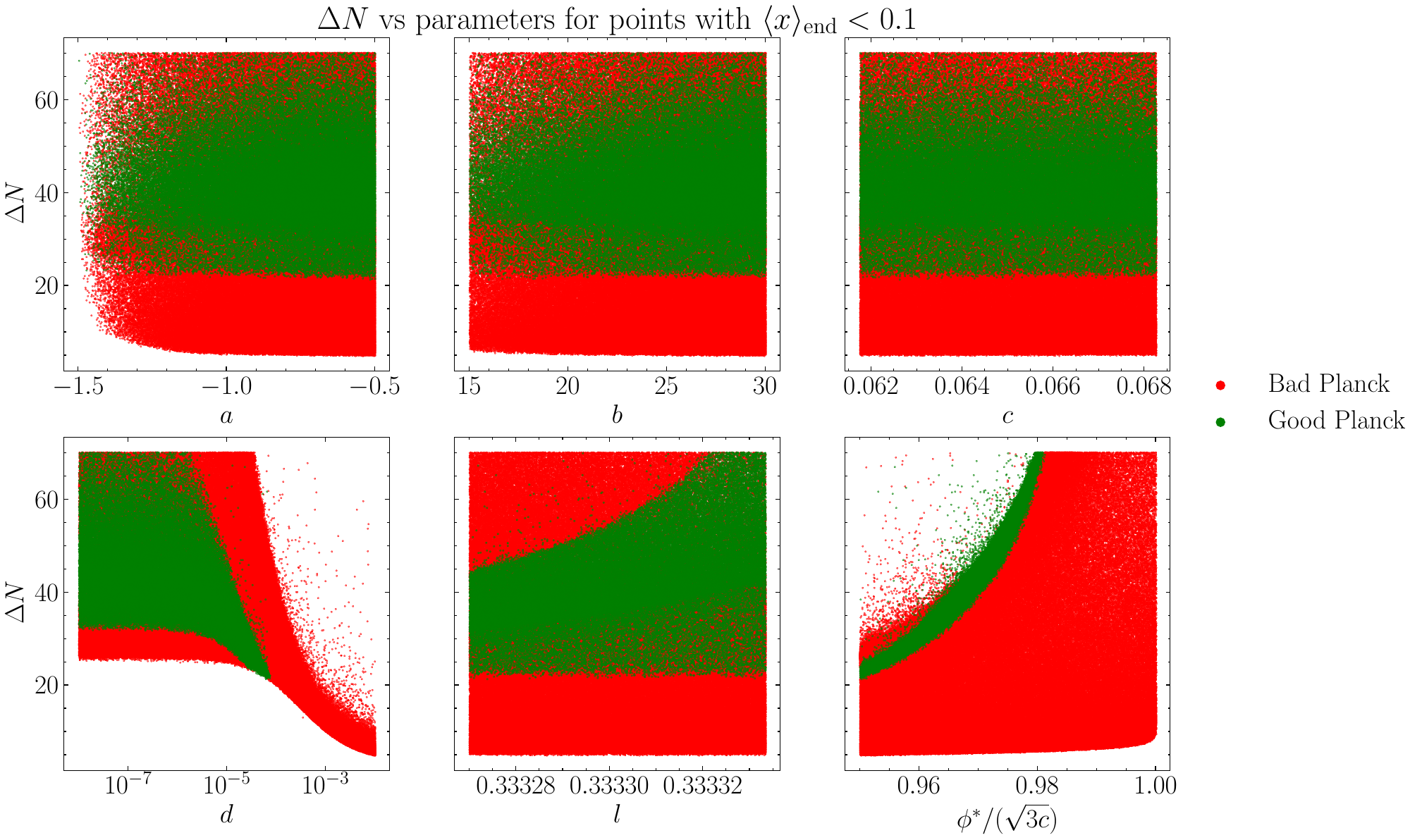}
    \caption{Number of e-folds, $\Delta N$ at the end of the numerical integration versus the parameters of the scalar potential. Green (Red) points agree (disagree) with Planck constraints~\cref{eq:planck_observables}.}
    \label{fig:n_final_vs_parameters}
\end{figure}

\begin{figure}[H]
    \centering
    \includegraphics[width=0.9\textwidth]{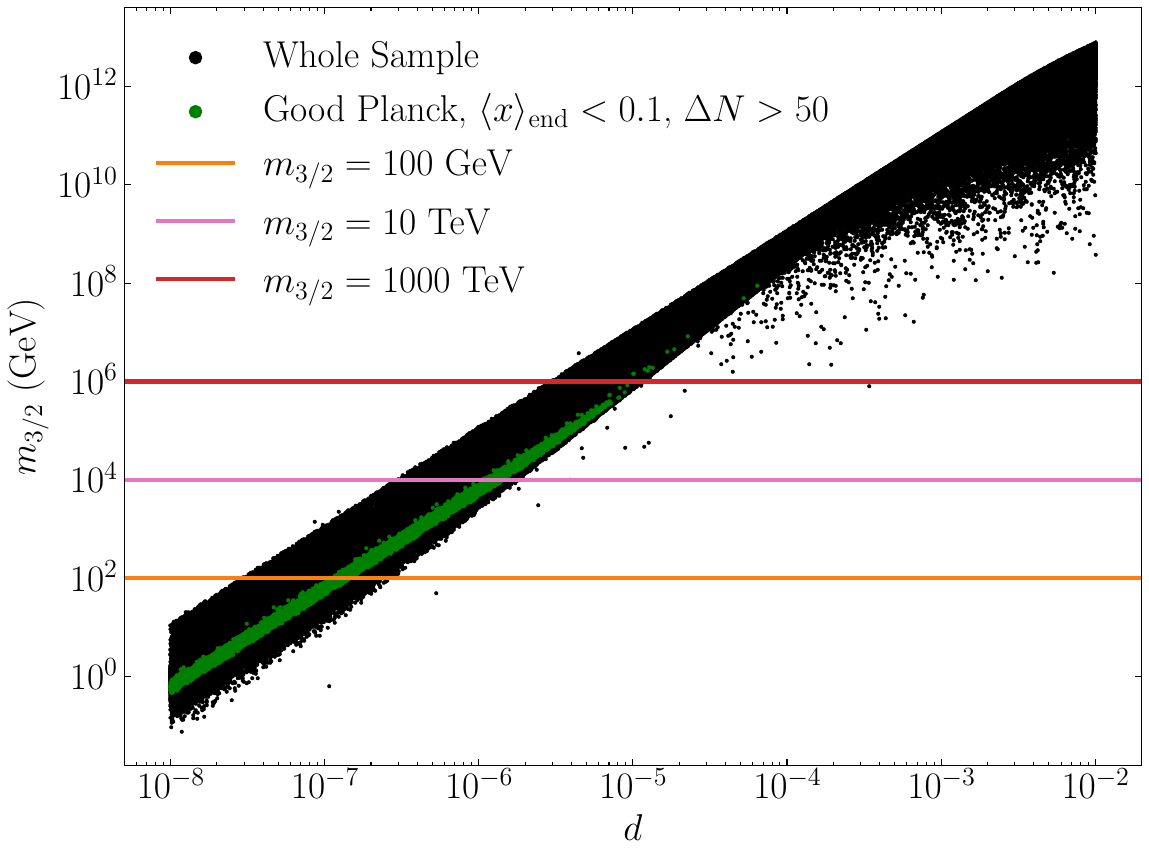}
    \caption{Gravitino mass (in GeV) as a function of $d$ (in units of $m_{Pl}$). Green points are in agreement with Planck constraints, and produce the required inflationary dynamics with the inflaton rolling down to the global minimum. Black points represent the whole sample. Horizontal lines are given for gravitino mass values of 100 GeV, 10 TeV, 1000 TeV.}
    \label{fig:m32_vs_d}
\end{figure}

Since we are interested in the cases where the inflaton reaches the global minimum, we now restrict to the case the points where $\langle x \rangle_{end} < 0.1$. For this subset of points, we show in~\cref{fig:n_final_vs_parameters} the number of e-folds, $\Delta N$, against the same parameters of the scalar potential. As expected, there are fewer points for smaller values of $a$ and $b$, as in this region the kink becomes as wall.
Focusing on the points that agree with the Planck satellite experiment, we see that there is a clear preference for $l \to 1/3$, which represents the Starobinsky limit of the ENO inflationary model. Additionally, we observe that $\phi^*$ needs to be fairly close to its upper bound for $\Delta N>50$, this is because for lower values of $\phi^*$ the inflaton starts very close to the inflexion of the potential providing a very short inflationary period. Finally, see how there is an upper bound on the parameter $d$ that sets the value of the Polonyi term. Since the Polonyi term is responsible for the non-gravitino mass, this in turn means that there is an upper bound on the values of the gravitino mass. In~\cref{fig:m32_vs_d} we show the possible values of the gravitino mass against $d$, and we observe that there is an upper bound of $\mathcal{O}(10^3)$ TeV on the gravitino mass for points with successful inflation. This is in agreement with our previous results from~\cite{Romao2017-pj}.

\begin{figure}[H]
    \centering
    \includegraphics[width=0.9\linewidth]{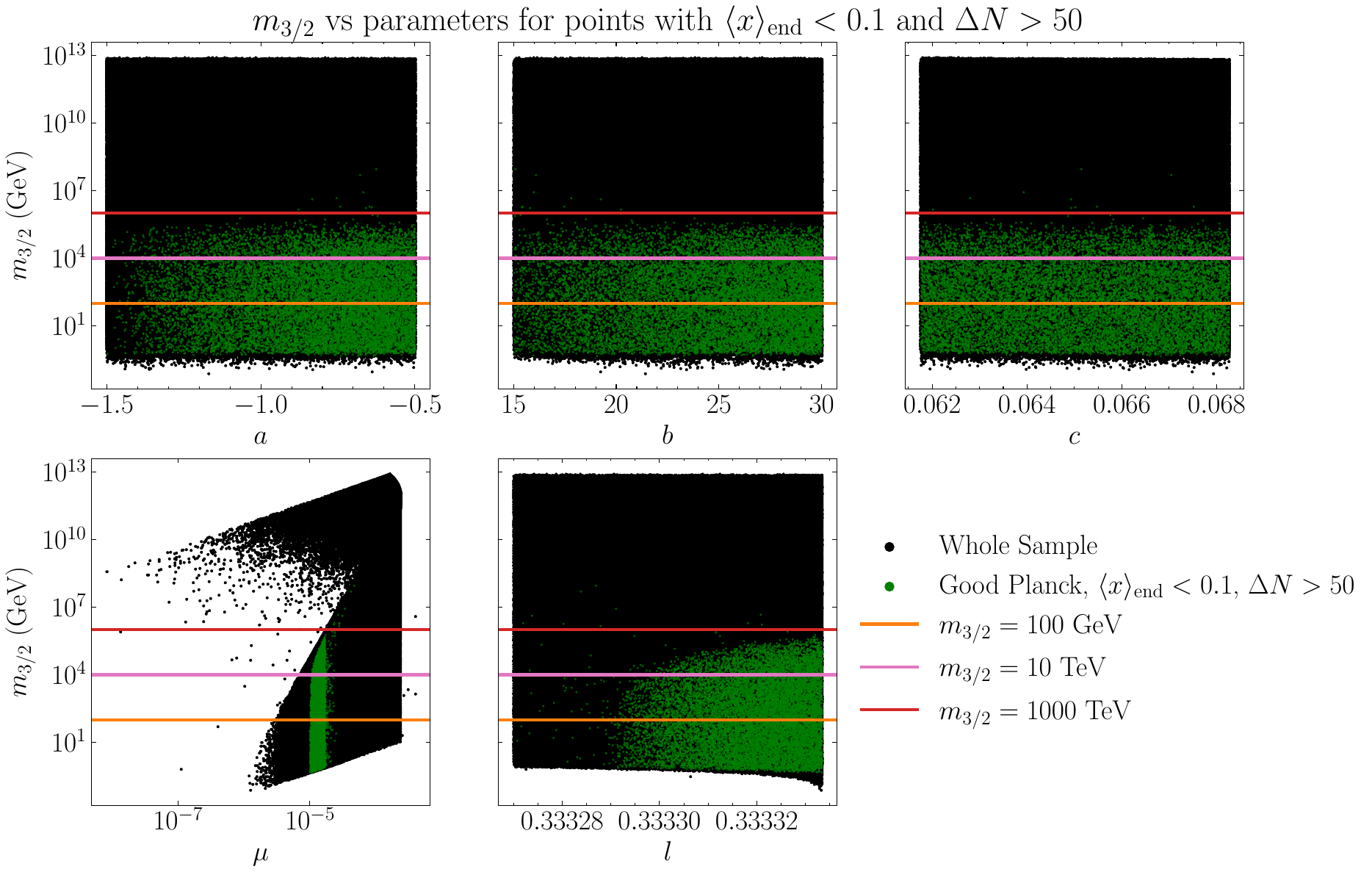}
    \caption{Gravitino mass (in GeV) versus (Top) \kahler\ potential and (Bottom) superpotential parameters. Green points agree with Planck constraints~\cref{eq:planck_observables}.}
    \label{fig:m32_vs_parameters}
\end{figure}

In~\cref{fig:m32_vs_parameters} we explore the sensitivity of the gravitino mass with regards the rest of the parameters. In the top row, we show the gravitino mass versus the \kahler\ potential parameters $a$, $b$, and $c$, and we observe that $m_{3/2}$ is uncorrelated to these. In the bottom row, we show $m_{3/2}$ versus the other two superpotential parameters $\mu$ and $l$. We observe that the gravitino mass is unrelated to $l$, but has a non-trivial dependence on $\mu$. However, in the region with valid points in green, we can see that the gravitino mass is uncorrelated with $\mu$, with $\mu \gtrsim 10^{-5}$ for all valid points.

\subsection{Exploratory Analysis of the Power Spectrum Peak\label{sec:exploratory_PR}}

So far we have only studied the parameter space points under Planck constraints and the requirements for the desirable inflationary dynamics. However, a crucial part of this work is to study the production of GW by an enhancement of the power spectrum originating by the presence of an inflexion point in the inflaton potential. As it has been discussed before~\cite{Nanopoulos:2020nnh, Stamou:2021qdk}, to fully study the peak of the power spectrum we need to go beyond the slow-roll approximation. However, this process is too computationally intensive to systematically study the parameter space. Instead, in this section we perform an initial analysis of the power spectrum using the slow-roll approximation,~\cref{eq:PRsr}, to identify potentially interesting points. Once a set of interesting points has been identified, we will compute the full numerical solution for the power spectrum in~\cref{sec:GW}.

\begin{figure}[H]
    \centering
    \includegraphics[width=1\textwidth]{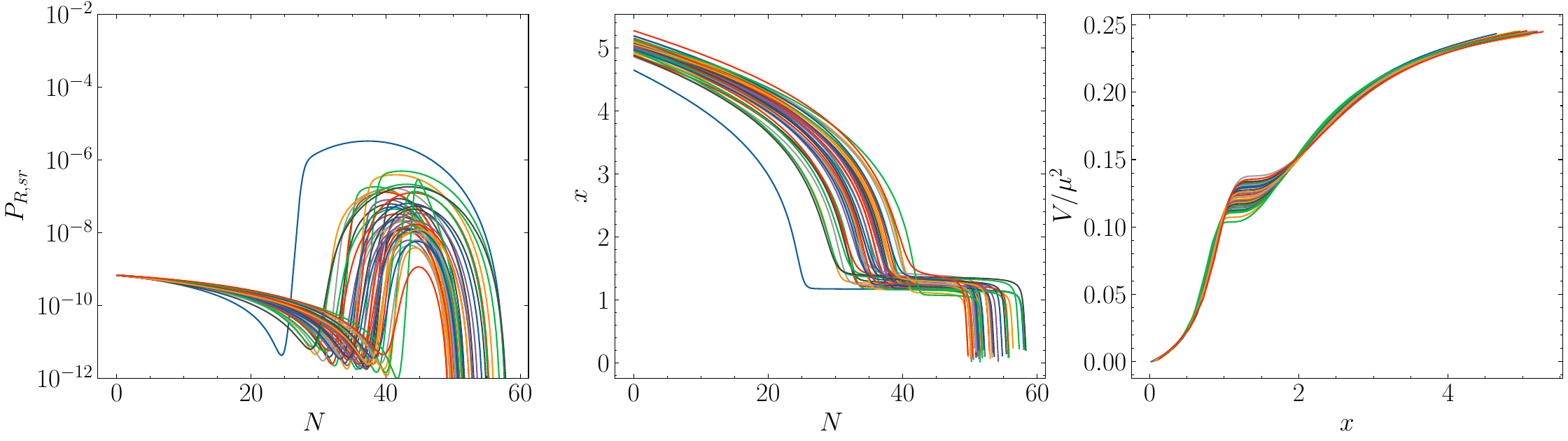}
    \caption{For the $46$ points passing all constraints we present, from left to right: The power spectrum in terms of the number of e-folds, the inflaton field value in terms of e-folds, and the inflaton potential in terms of $x$. However the enhancement of the power spectrum in these examples is not sufficient to produce observable GW.}
    \label{fig:naive_promising_points}
\end{figure}

We first look at the points that lead to good inflationary dynamics and observables, i.e,
\begin{itemize}
    \item Agree with Planck constraints
    \item $50 \leq \Delta N \leq 60$
    \item $\langle x \rangle_{end} < 0.1 $\ .
\end{itemize}
In addition, we further restrict to points where the (slow-roll) power spectrum, $P_{R,s.r.}$, has a maximum when the inflaton rolls through the inflexion point. Defining $\tilde N$ as the e-fold time at which the (slow-roll) has a maximum, we then restrict it to be around $ 30 \lesssim \tilde N \lesssim 45$. Out of $10^6$ points, only $46$ satisfy these conditions. In~\cref{fig:naive_promising_points} we show the power spectrum, $P_{R,s.r.}$, the inflaton field value, $x$, in terms of the e-fold time, $N$, and the potential for these $46$ points.

Although all the $46$ points pass the constraints listed above, they are actually not good candidates for GW production. The reason being that $P_{R,s.r.}$ is not enhanced enough, as we would need $P_{R,s.r.}$ to reach around $10^{-4}$ to produce observable gravitational waves. In principle, this could be obtained by tweaking the value of the parameter $b$, which has been observed to produced the desired enhancement with significant level of fine-tuning~\cite{Stamou:2021qdk}. The tuning, however, does not work for these points for the following reason. Enhancing the $P_{R,s.r.}$ peak requires $x$ to slow down as it passes the plateau even further. Looking at the middle and right plots in~\cref{fig:naive_promising_points}, this requires $x$ to spend \emph{more time} rolling through the lower plateau. This can be achieved by slightly decreasing the value of $b$, as this enhances kink. However, when tuning $b$ for these points we find that $\Delta N$ can easily go beyond $60$, producing \emph{too much inflation}, while the $P_{R,s.r.}$ peak never raises significantly. We show this explicit for one of the four points where $\Delta N \simeq 50$, given by the parameters
\begin{align}
    \{a,b,c,d,l,\phi^*\} = \{-0.827, 19.0177, 0.0666, 5.35\times 10^{-6}, 0.3333, 0.432\}\label{eq:naive_point_1} \ ,
\end{align}
and where we keep more significant digits for $b$ and $l$ parameters to which the potential and the power spectrum are especially sensitive to. We now tune the $b$ parameter by reducing it by a small amount parametrised by $\delta b$ such that
\begin{align}
    b \to b \times(1 - \delta b) \ ,
\end{align}
where we scan the for the possible values $\delta b \in 4\times[10^{-5},10^{-4}]$.

\begin{figure}[H]
    \centering
    \includegraphics[width=1\textwidth]{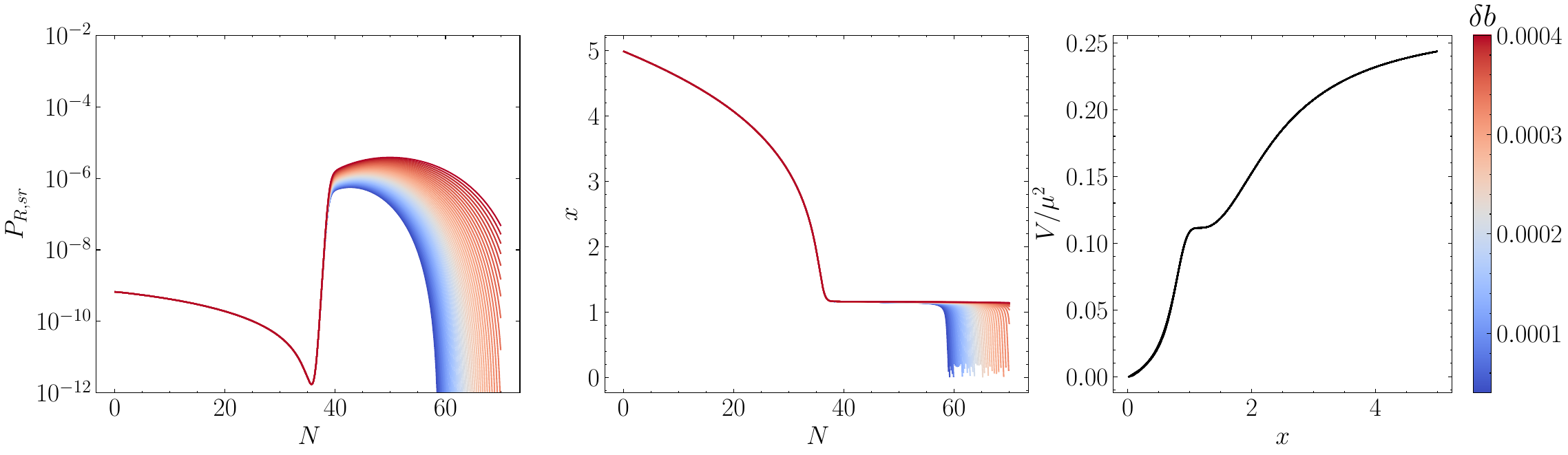}
    \caption{For the point in~\cref{eq:naive_point_1} we present, from left to right and for various values of $b$: The power spectrum in terms of the number of e-folds, the inflaton field value in terms of e-folds, the inflaton potential. As before the enhancement of the power spectrum in these examples is not sufficient to produce observable GW.}
    \label{fig:naive_promising_point_1}
\end{figure}

In~\cref{fig:naive_promising_point_1} we show the changes as we tune the value of $b$. In the leftmost plot, we see that even though $P_{R,s.r.}$ can be enhanced by an order of magnitude, it fails to be significantly increased. In the middle plot, we observe how the number of e-folds increase beyond $60$ as we decrease $b$. This is easily understood as by decreasing $b$ we are slowing down the inflaton field at the plateau, prolonging inflation. In the limit that we significantly decrease $b$, we can see how the inflaton can get stuck at the plateau, as $x$ does not roll down to the bottom of the potential. Finally, on the rightmost plot we see that changes to $b$ of the order of the tuning, $\mathcal{O}(10^{-4})$, has no visible effect on the shape of the potential, even thought it has a significant impact on the $P_{R,s.r.}$ and the number of e-folds. Similar conclusions can be drawn for the remaining 45 points that pass all of the constraints.

\begin{figure}[H]
    \centering
    \includegraphics[width=1\textwidth]{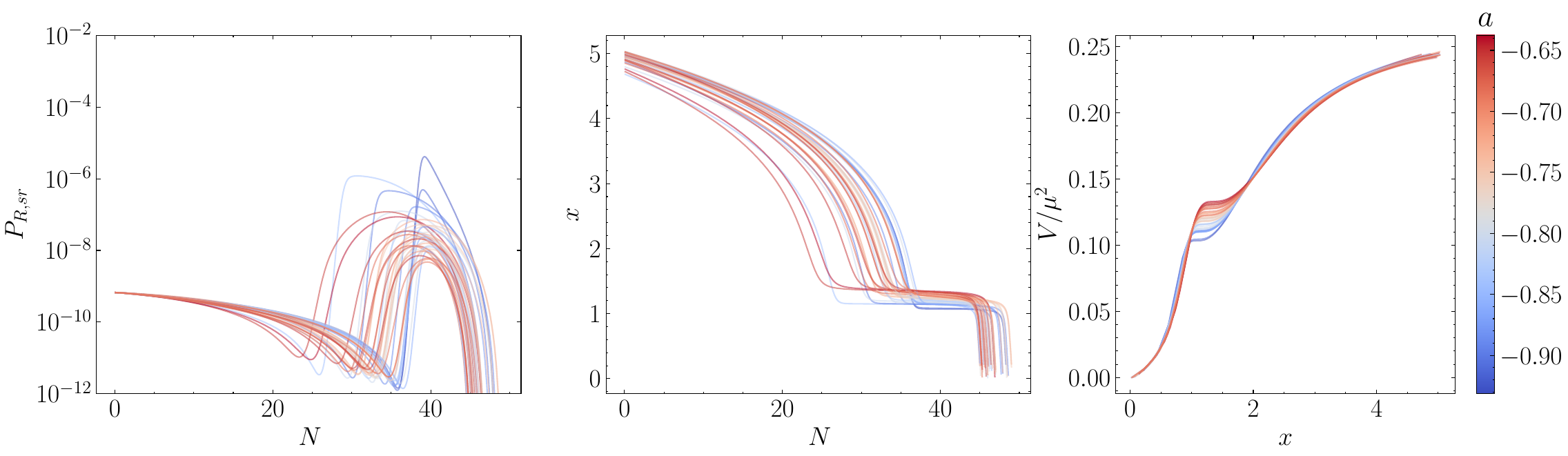}
    \caption{Promising points that agree with Planck, but with shorter inflationary period of $45 \leq \Delta N \leq 50$ and earlier $P_R$ peak ($30 < \tilde N <40$). Gradient represents the value of $a$ of each point. From left to right: The power spectrum in terms of the number of e-folds, the inflaton field value in terms of e-folds, the inflaton potential. However the enhancement of the power spectrum in these examples is still insufficient to produce observable GW.}
    \label{fig:promising_points_vs_a}
\end{figure}

\begin{figure}[H]
    \centering
    \includegraphics[width=1\textwidth]{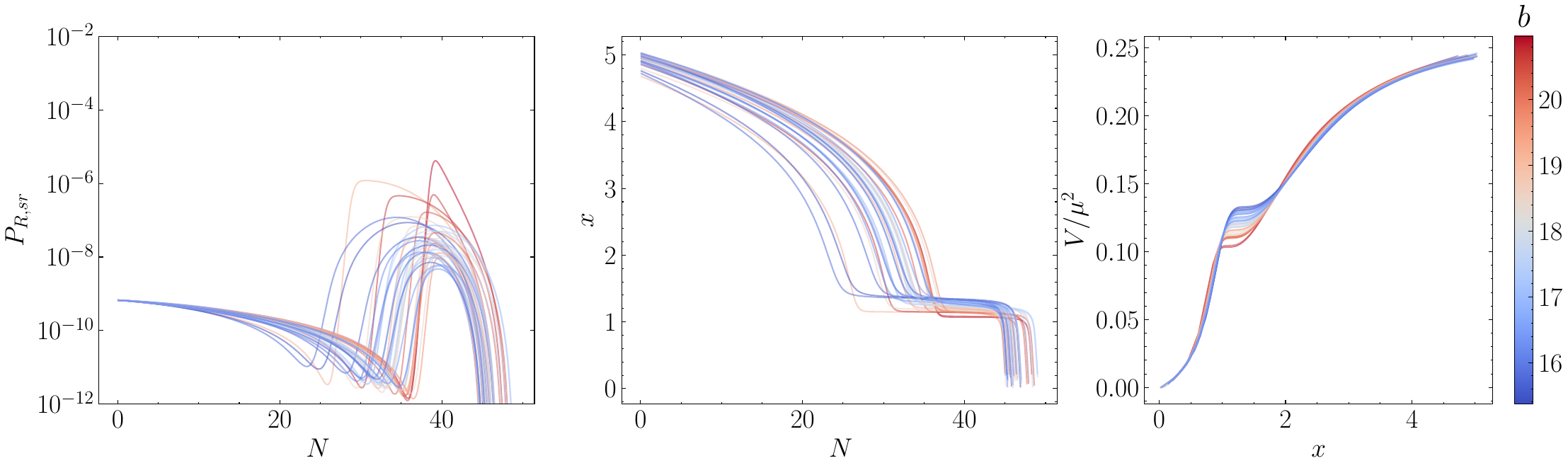}
    \caption{Promising points that agree with Planck, but with shorter inflationary period of $45 \leq \Delta N \leq 50$ and earlier $P_R$ peak ($30 < \tilde N <40$). Gradient represents the value of $b$ of each point. From left to right: The power spectrum in terms of the number of e-folds, the inflaton field value in terms of e-folds, the inflaton potential. Again the enhancement of the power spectrum in these examples is not sufficient to produce observable GW.}
    \label{fig:promising_points_vs_b}
\end{figure}

\begin{figure}[H]
    \centering
    \includegraphics[width=0.4\textwidth]{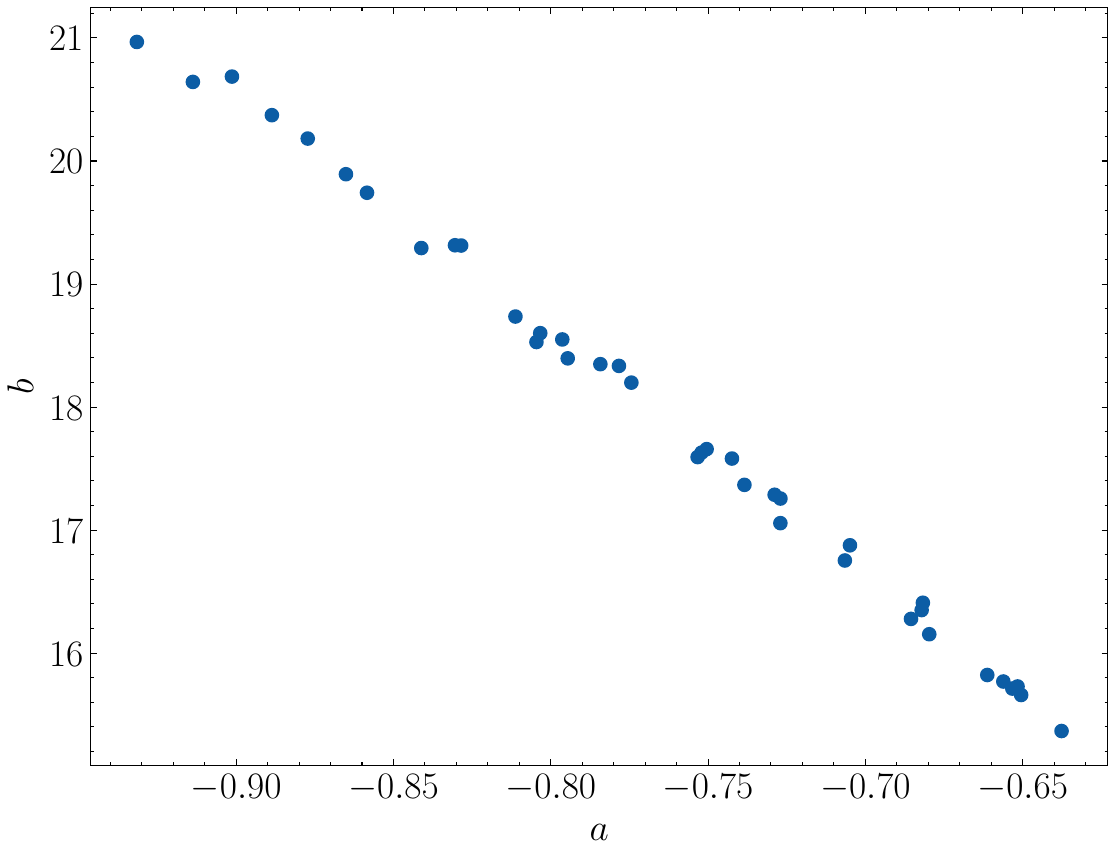}
    \caption{$(a,b)$ scatter plot for the points that agree with Planck, but with shorter inflationary period of $45 \leq \Delta N \leq 50$ and earlier $P_R$ peak ($30 < \tilde N <40$).}
    \label{fig:promising_points_a_vs_b}
\end{figure}

The previous discussions opens up for the possibility that there may be points that have not passed all the required constraints, but might be tunable into promising points. We now focus on points that have a good agreement with Planck data, but that the inflaton rolls down to the global minimum earlier, also producing an earlier $P_{R,s.r.}$ peak, i.e.
\begin{align}
    45 \leq \Delta N \leq 50 \label{eq:early_Delta_N} \\
    30 < \tilde N <40 \ , \label{eq:early_N_tilde}
\end{align}
as we expect to be able to \emph{delay} both quantities by tuning $b$ accordingly. Out of the million sampled points, $38$ points satisfy these requirements. In~\cref{fig:promising_points_vs_a,fig:promising_points_vs_b} we present the power spectrum, the inflaton field values, and the potential for these points, and where the colour of the plots is given by the value of the parameters $a$ and $b$, respectively. In accordance to the discussion in~\cref{sec:V}, we observe that lower values of $b$ lead to the plateau (arising from the inflexion point) happens \emph{earlier}, i.e. for smaller values of both the e-fold time, $N$, and the value of the inflaton field, $x$.
Perhaps more surprisingly, we also see the same trend for $a$, which is a phenomenon not supported by the earlier discussion in~\cref{sec:V}. In fact, in~\cref{fig:promising_points_a_vs_b} we show that for these points\footnote{In fact, the same is observed for the points that pass all the constraints shown in~\cref{fig:naive_promising_points}, but this is not shown to declutter the discussion.}, $a$ and $b$ are highly anti-correlated. In light of the discussion in~\cref{sec:V}, this can be understood as follows. Decreasing $b$ will produce an earlier plateau around the inflexion point, but it might also turn the plateau into a wall. On the other hand, decreasing $a$ can mitigate the appearance of the wall after the inflexion point, and to an even shorten the plateau\footnote{We have also performed an equivalent analysis for the parameter $c$, but there was no correlation.}.

\begin{figure}[H]
    \centering
    \includegraphics[width=1\textwidth]{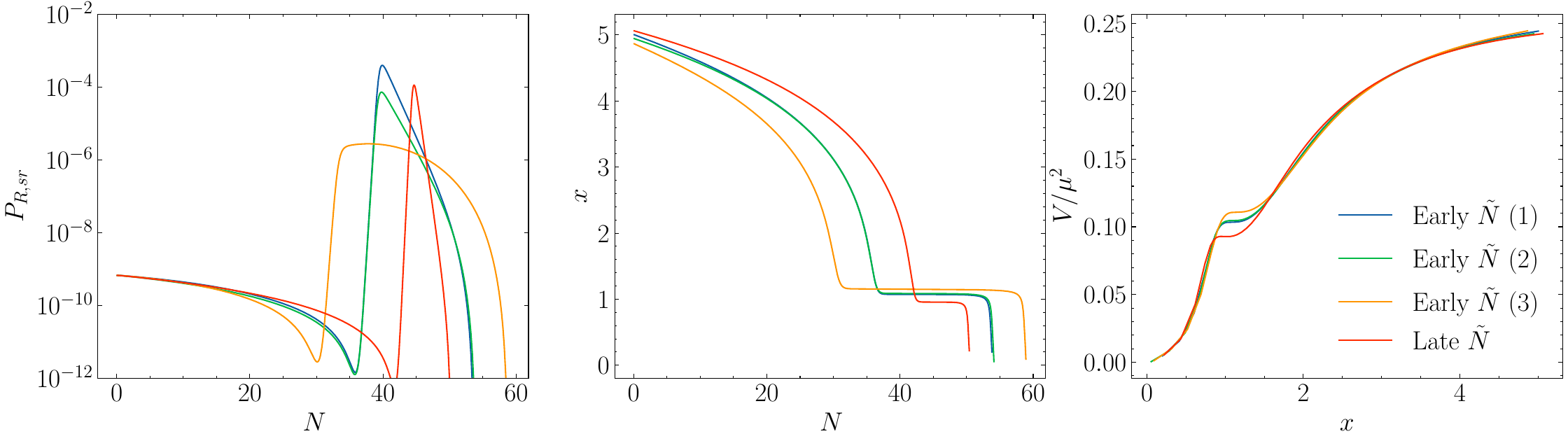}
    \caption{Promising points from~\cref{tab:tweaked_good_points} with $b$ tuned. In this case the enhancement of the power spectrum in these examples is sufficient to produce observable GW.}
    \label{fig:tweaking_good_points}
\end{figure}

Another feature that is noticeable from~\cref{fig:promising_points_vs_a,fig:promising_points_vs_b} is that for larger values of $a$ the power spectrum has a wider and rounder peak, similar to what can be seen in~\cref{fig:naive_promising_point_1} and for most of the points in~\cref{fig:naive_promising_points}. As it was studied in~\cref{fig:naive_promising_point_1}, these are unlikely to produced a high enough power spectrum enhancement. We therefore focus on the points where the power spectrum has a sharp enhancement, which happens for $a \leq - 0.9$. This cut produces only three points, which we will denote Early $\tilde N$ points. We then proceed to tune $b$ in order to maximise $P_{R,s.r.}$ while keeping $\Delta N \lesssim 60$ and avoiding $x$ getting stuck at the plateau. For completeness of our analysis, we will also include a point with a slightly later $\tilde N$, i.e. a point that initially (prior to tuning) respects~\cref{eq:early_Delta_N} but not~\cref{eq:early_N_tilde}. The four points identified by the random scan as showing promising features for gravitational waves production to be studied in more detail in~\cref{sec:GW} are presented in~\cref{tab:tweaked_good_points} and in~\cref{fig:tweaking_good_points}.
\begin{table}[H]
    \centerline{
        \begin{tabular}{lcccccccc}
            {}                   & $a$      & $b$         & $c$      & $d$                  & $l$       & $\phi^*$ & $\delta b$ & $b\times(1-\delta b)$ \\
            \hline
            \hline
            Early $\tilde N$ (1) & $-0.932$ & $20.967463$ & $0.0671$ & $8.56\times 10^{-6}$ & $0.33331$ & $0.434$  & $0.000056$ & $20.966291$           \\
            Early $\tilde N$ (2) & $-0.914$ & $20.642496$ & $0.0671$ & $9.26\times 10^{-7}$ & $0.33330$ & $0.433$  & $0.000206$ & $20.638251$           \\
            Early $\tilde N$ (3) & $-0.901$ & $20.686070$ & $0.0618$ & $9.64\times 10^{-6}$ & $0.33328$ & $0.415$  & $0.000475$ & $20.676242$           \\
            Late $\tilde N$      & $-1.180$ & $25.427327$ & $0.0645$ & $2.02\times 10^{-6}$ & $0.33333$ & $0.426$  & $0.000007$ & $25.427144$           \\
        \end{tabular}
    }
    \caption{Promising points with $b$ tuned in order to maximise $P_{R,s.r.}$ and to achieve $\Delta N\gtrsim 50$. Given the level of tuning associated with $b$ and the sensitivity of the shape of potential to $l$, we express these parameters with more significant digits than $a$, $c$, $d$, and $\phi^*$.}
    \label{tab:tweaked_good_points}
\end{table}

In~\cref{tab:tweaked_points_observables} we present the observables $n_s$, $r$, and $m_{3/2}$ for the tuned points listed in~\cref{tab:tweaked_good_points}, where we highlight they all agree with Planck satellite constraints from~\cref{eq:planck_observables}. On the other hand, the gravitino mass values are within what we expected from the discussion earlier in this section. It is worth noticing that the observables in this table changed at most $\mathcal{O}(10^{-5})$ as we tuned $b$, which is far below the number of significant digits presented. This reinforces the idea that tuning $b$ will have a considerable impact on the power spectrum peak and number of e-folds, but not on the remainder of the observables, including the gravitino mass.
\begin{table}[H]
    \centering
    \begin{tabular}{lccc}
        {}                   & $n_s$    & $r$     & $m_{3/2}$    (GeV)  \\
        \hline
        \hline
        Early $\tilde N$ (1) & $0.9677$ & $0.011$ & $7.24\times 10^{5}$ \\
        Early $\tilde N$ (2) & $0.9576$ & $0.009$ & $7.74\times 10^{3}$ \\
        Early $\tilde N$ (3) & $0.9674$ & $0.017$ & $1.31\times 10^{6}$ \\
        Late  $\tilde N$     & $0.9586$ & $0.007$ & $3.35\times 10^{4}$ \\
    \end{tabular}
    \caption{Observables for the tuned points from~\cref{tab:tweaked_good_points}.}
    \label{tab:tweaked_points_observables}
\end{table}

\section{Full Numerical Solution and Gravitational Waves Production\label{sec:PR_and_GW}}

In the previous section we have performed a thorough parameter space scan where we identified points that produce the desired inflationary observables, and that could be tuned such that the slow-roll approximated power spectrum was enhanced, while ensuring that the inflaton rolls down to the global minimum after around $50$ e-folds. However, it has been pointed out~\cite{Nanopoulos:2020nnh,Stamou:2021qdk,Spanos:2021hpk,Spanos:2022euu} that the slow-roll approximated power spectrum in~\cref{eq:PRsr} fails to capture the true power spectrum, where the peak can be both higher and be shifted. This is easily understood as the dynamics around the plateau break the slow-roll approximation, since the inflaton field arrives at the plateau with high velocity, invalidating $\epsilon_V \ll 1$, and is met with an increase in deceleration, invalidating $|\eta_V| \ll 1$. Once the full numerical solution for the power spectrum is found, we can then compute the present day gravitational wave density spectrum.

\subsection{Full Numerical Solution for Power Spectrum\label{sec:PR}}

We now focus on accurately computing the power spectrum. The procedure presented here is detailed in~\cite{Ringeval:2007am}, and it has been applied in~\cite{Spanos:2021hpk,Spanos:2022euu}. For a given mode of comoving wavelength $k$, the associated power spectrum is given by
\begin{align}
    P_R & = \frac{k^3}{2 \pi^2} |R_k|^2    \label{eq:PRfull} \\
    R_k & = \Psi + \frac{\delta x}{dx/dN} \ ,
\end{align}
where $\Psi$ is the Bardeen potential and $\delta x$ is the inflaton fluctuations field\footnote{Strictly speaking, the $\delta x$ refers to the mode $k$, i.e. $\delta x_k$, as we are studying the contribution of each mode to the power spectrum. Here we follow the slightly simplified notation also found elsewhere in the literature.}. The fields $\Psi$ and $\delta x$ are obtained by evolving their respective equation of motion
\begin{align}
    \frac{d^2 \delta x}{d N^2} & =-\left(3 - \frac{1}{2}\left(\frac{d x}{d N}\right)^2\right) \frac{d \delta x}{d N} - \frac{1}{H^2} \partial_{x,x} V \delta x - \frac{k^2}{a^2 H^2} \delta x + 4 \frac{d \Psi}{d N}\frac{d x}{d N} - \frac{2 \Psi}{H^2}\partial_x V \label{eq:delta_x_EOM} \\
    \frac{d^2 \Psi}{d N^2}     & = - \left(7 - \frac{1}{2}\left(\frac{d x}{d N}\right)^2\right) \frac{d \Psi}{d N} - \left(2 \frac{V}{H^2}+\frac{k^2}{a^2 H^2}\right)\Psi - \frac{1}{H^2}\partial_{x,x}V \delta x \ . \label{eq:Psi_EOM}
\end{align}

The initial conditions are set at subhorizon scales, $k\gg a H$, such that we have the so-called Bunch-Davies vacuum. For both $\delta x$ and $\Psi$, as well as their derivatives, the initial conditions can be written solely as dependent on the the comoving wavelength, $k$, and quantities that can be derived from the background solution for $x$:
\begin{align}
    (\delta x)_{i.c.}                          & = \frac{1}{\sqrt{2 k}}\frac{1}{a_{i.c.}}       \label{eq:delta_x_ic}                                                                                                                                                                                                                                    \\
    \left(\frac{d \delta x}{d N}\right)_{i.c.} & = - \frac{1}{a_{i.c.} \sqrt{2 k}}\left(1+i \frac{k}{a_{i.c.} H_{i.c.}}\right)  \label{eq:d_delta_x_ic}                                                                                                                                                                                                  \\
    (\Psi)_{i.c.}                              & = \frac{1}{2\left(\epsilon_{H,i.c.}-\frac{k^2}{a^2_{i.c.} H^2_{i.c.}}\right)}\left[\left(\frac{d x}{d N}\right)_{i.c.}\left(\frac{d \delta x}{d N}\right)_{i.c.} + (\delta x)_{i.c.} \left(3\left(\frac{d x}{d N}\right)_{i.c.}+\frac{1}{H_{i.c.}}(\partial_x V)_{i.c.}\right)\right] \label{eq:Psi_ic} \\
    \left(\frac{d\Psi}{d N}\right)_{i.c.}      & = \frac{1}{2} \left(\frac{dx}{dN}\right)_{i.c.} (\delta x)_{i.c.} - \Psi_{i.c.} \ , \label{eq:d_Psi_ic}
\end{align}
where every quantity with subscript $i.c.$ is computed at an initial conditions e-fold time, $N_{i.c.}$.

The process to obtain the power spectrum as a function of the comoving wavelength of the modes, $P_R=P_R(k)$, is highlighted in~\cite{Ringeval:2007am} and we summarise it now.
\begin{enumerate}
    \item The background solution for $x$,~\cref{eq:EOM_x} is integrated from $N^*=0$ until $N_{end}= \Delta N$, from which we obtain $x(N)$ and $d x/dN (N)$.
    \item For each comoving wavelength $k'$, we can find the e-fold time at which it crosses the Hubble radius as $k' = a(N') H(N')$.
    \item The initial conditions for mode $k'$ are set at a time $N_{i.c.}$ earlier than $N'$, ascribed by the relation~\cite{Ringeval:2007am}
          \begin{equation}
              k' = C_q a(N_{i.c.})H(N_{i.c.}) \ ,
          \end{equation}
          where we set $C_q=100$. The value of $N_{i.c.}$ is obtained by producing an interpolation map $N(k)$ obtained by inverting the function $k(N)=a(N)H(N)$\footnote{There are two subtle details to refer in this step. The first one, is that using the background solution, $x$, and $k^*=a(N^*) H(N^*)$ we can obtain $a(N^*)=a(0)$, from which we can find $a(N)$ for any other e-fold time $a(N)=a(0) e^{N}$. The second detail is more nuanced: for the pivot scale measured by the CMB, $k^*=0.05\ Mpc^{-1}$, we have $N'=N*=0>N_{i.c.}$ and therefore the initial conditions have to be set for a negative $N_{i.c.}$. To workaround this, the background solution for $x$ is in fact evolved from $N=-3$ until $\Delta N$. Effectively this adds a few more e-folds to inflation prior to the CMB crossing the Hubble radius, while keeping the pivot scale well defined and identified with the Planck experiment.}. All quantities in the initial conditions,~\cref{eq:delta_x_ic,eq:d_delta_x_ic,eq:Psi_ic,eq:d_Psi_ic}, can be computed using the background solution for $x$ and its derivative.
    \item The equations of motion~\cref{eq:EOM_x,eq:delta_x_EOM,eq:Psi_EOM} are then integrated jointly from $N_{i.c.}$ until $N_{end}=\Delta N$ in order to provide to obtain the present day power spectrum associated with the mode of wavelength $k$, and~\cref{eq:PRfull} is finally evaluated.
    \item The process is repeated for values of $k'\in[k^*, k(\Delta N)]$, in logarithm steps. This produces a list of tuples $(k,P_R)$ from which an interpolating function $P(k)$ can be obtained.
\end{enumerate}

Using the steps above, we were able to reproduce the power spectra in~\cite{Nanopoulos:2020nnh,Stamou:2021qdk,Spanos:2022euu}\footnote{Our results slightly disagree some of theirs, but this is due to a difference in the final form of the scalar potential already discussed in~\cref{sec:V} and this was resolved after contacting the authors.}
\begin{figure}[H]
    \centering
    \begin{subfigure}{0.4\textwidth}
        \includegraphics[width=\linewidth]{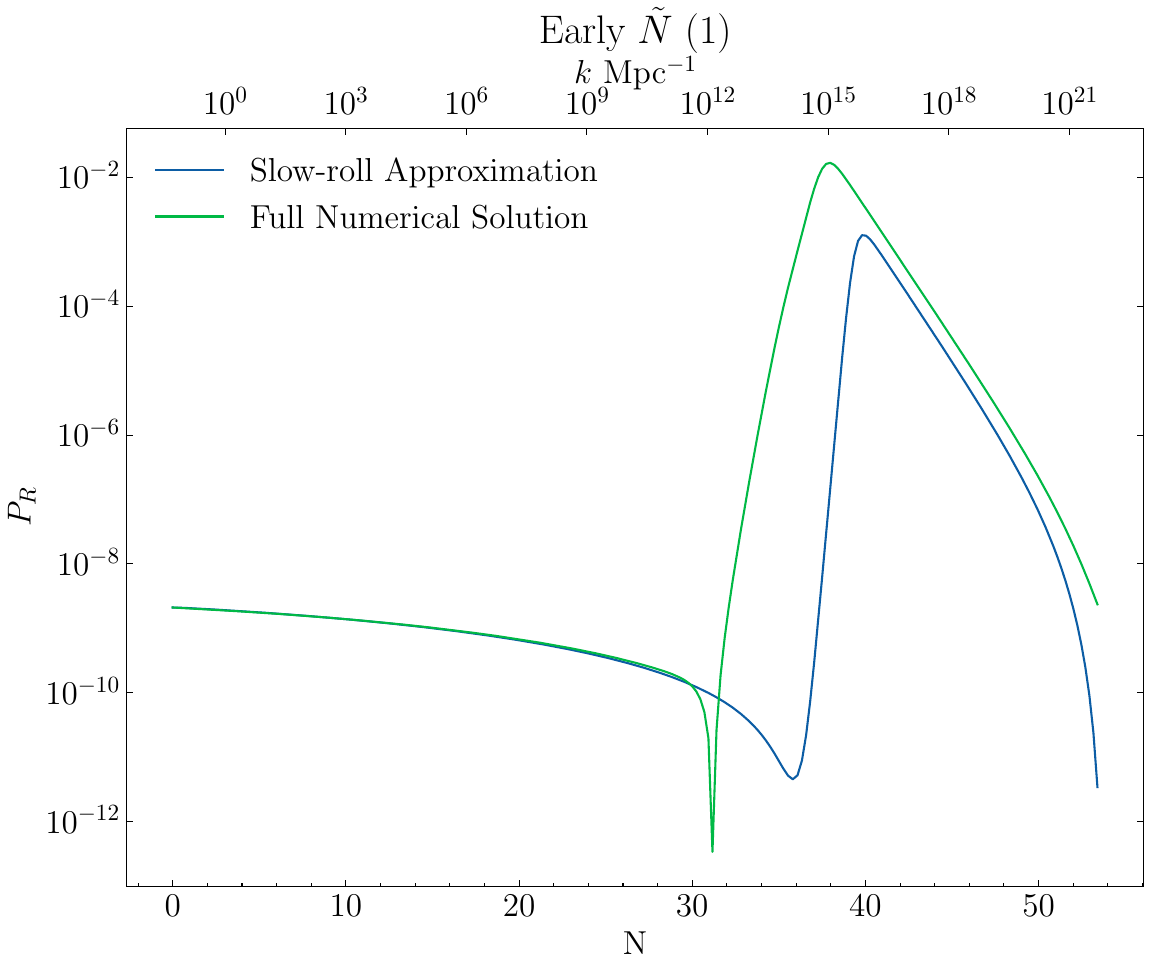}
    \end{subfigure}
    \begin{subfigure}{0.4\textwidth}
        \includegraphics[width=\linewidth]{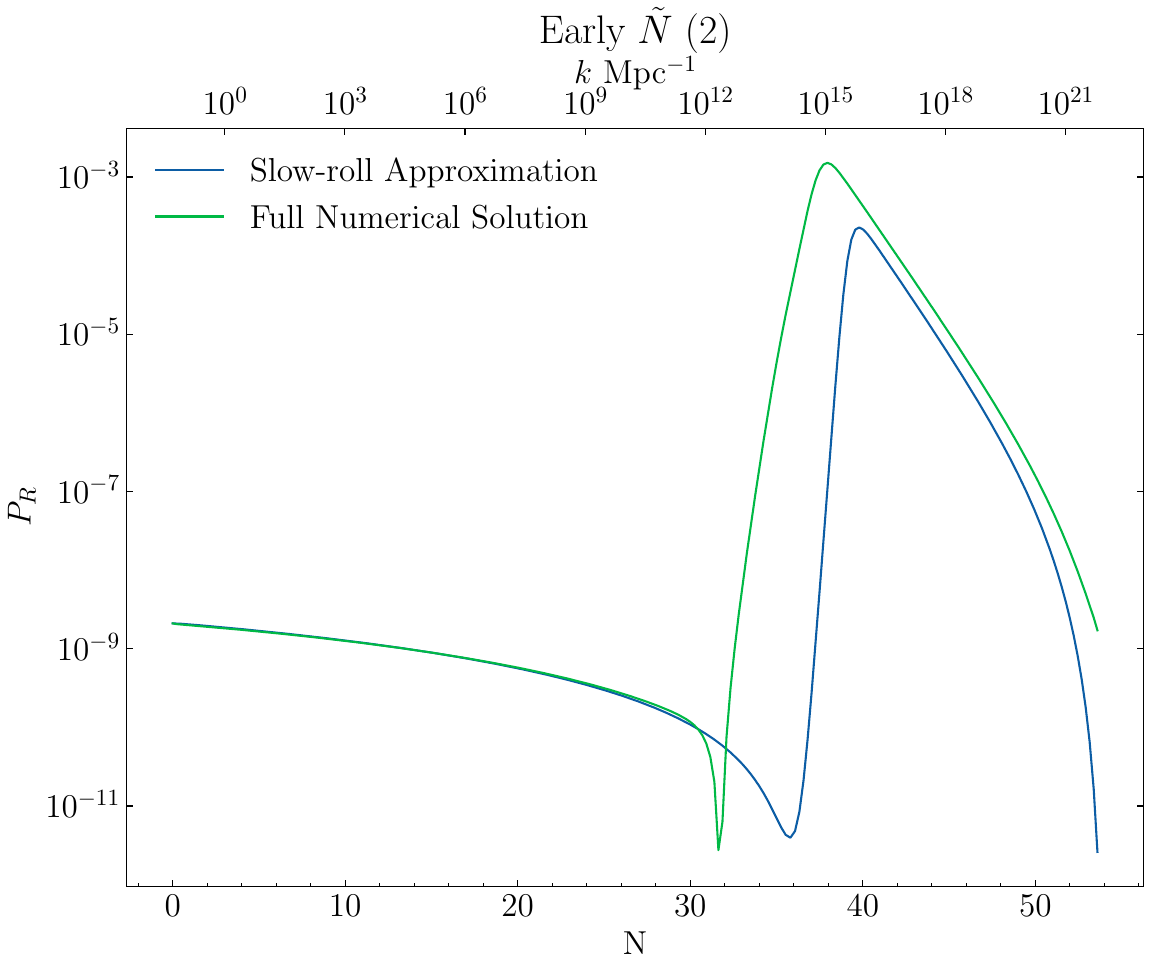}
    \end{subfigure}
    \begin{subfigure}{0.4\textwidth}
        \includegraphics[width=\linewidth]{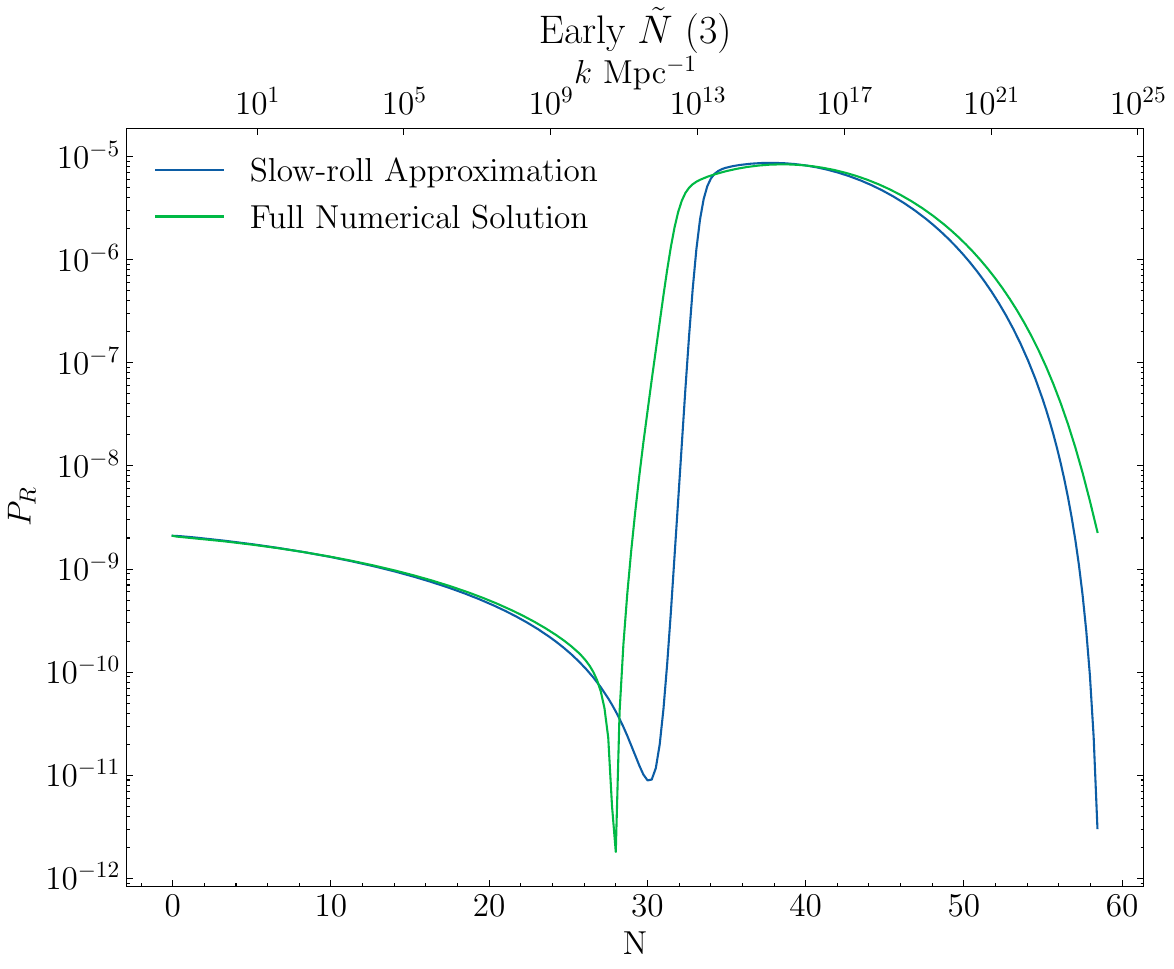}
    \end{subfigure}
    \begin{subfigure}{0.4\textwidth}
        \includegraphics[width=\linewidth]{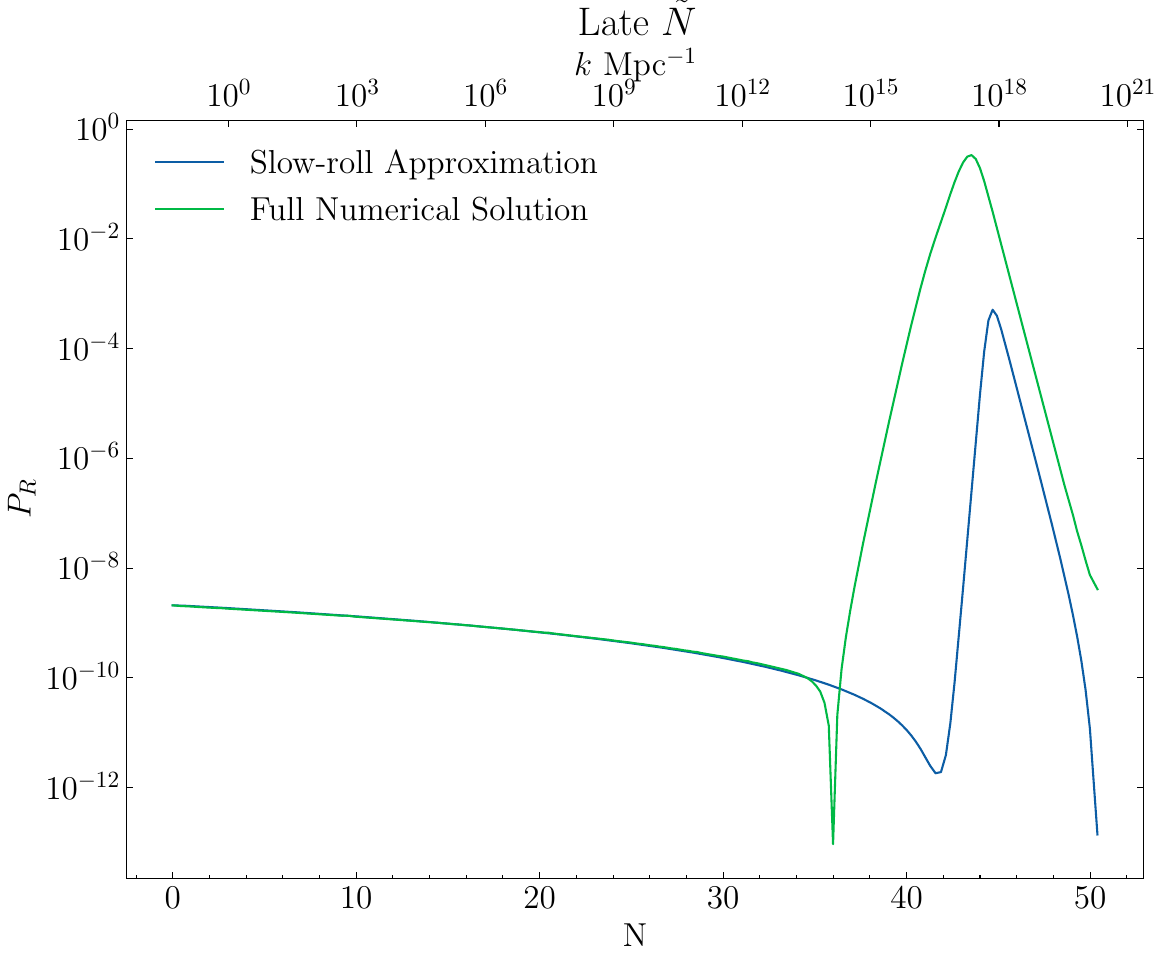}
    \end{subfigure}
    \caption{Comparison between the slow-roll approximation power spectrum, c.f.~\cref{eq:PRsr}, against the full numerical solution power spectrum, c.f.~\cref{eq:PRfull}. The points shown are those identified in~\cref{sec:scan} and are listed in~\cref{tab:tweaked_good_points}.}
    \label{fig:PR_vs_PR_SR_multiple_points}
\end{figure}
In~\cref{fig:PR_vs_PR_SR_multiple_points} we compare the power spectrum obtained by computing the full numerical solution of the equations of motion against the one provided by the slow-roll approximation for the points in~\cref{tab:tweaked_good_points}. A few things are worth noticing at this stage. Firstly, both methodologies agree for early stages of inflation, from $N\simeq N^*$ ($k\simeq k^*=0.05$ Mpc$^{-1}$) up until the cusp of the high plateau at $N\simeq 30$, which is the regime when the slow-roll approximation is intended to be used. Secondly, we notice that for points where the power spectrum is sharply enhanced by the plateau, it gets further amplified by at least a few orders of magnitude, whereas for the remainder points there is no discernable amplification. Thirdly, the power spectrum peak is shifted to the left, i.e. to happen earlier. Lastly, given that, in human units,
\begin{align}
    1\ \text{Mpc} ^{-1} = 0.97154 \times 10^{-14}\ \text{Hz},
    \label{eq:k_to_f}
\end{align}
we expect, to first approximation, that for earlier peaks the produced gravitational waves to be produced for frequencies $\mathcal{O}(0.1-1)$ Hz, while for later peaks the frequencies will be around $\mathcal{O}(100-1000)$ Hz.

\subsection{Gravitational Waves Production\label{sec:GW}}

With the power spectrum obtained from the previous section, we have now all the ingredients to compute the present day gravitational wave density. Following~\cite{Espinosa:2018eve,Spanos:2022euu}, the present day density gravitational wave with wave number $k$ is given by\footnote{For numerical purposes, as the integration in $v$ spans several orders of magnitude over the argument of $P_R$, we performed the integration over $\tilde v=\log v$ with the appropriate jacobian.}
\begin{align}
    h^2 \Omega_{GW}(k) = \frac{\Omega_r}{36}\int_0^{1/\sqrt3}du \int_{1/\sqrt{3}}^\infty dv \left[\frac{(u^2-1/3)(v^2-1/3)}{u^2-v^2}\right]^2 P_R(k X) P_R(k Y) (I^c_2+I^2_s) \ ,
    \label{eq:omegagw}
\end{align}
where $\Omega_r\simeq 5.4 \times 10^{-5}$ is the present day radiation density,
\begin{align}
    X & =\frac{\sqrt{3}}{2}(u+v)      \\
    Y & =\frac{\sqrt{3}}{2}(-u+v) \ , \\
\end{align}
and we use the analytical approximation for the functions $I_c$ and $I_s$ (see~\cite{Espinosa:2018eve} for more details)
\begin{align}
    I_c & = -36 \pi \frac{(u^2+v^2-2)^2}{(-u^2+v^2)^3} \Theta(v-u)                                                                     \\
    I_s & = -36 \frac{(u^2+v^2-2)^2}{(u^2-v^2)^2}\left[\frac{(u^2+v^2-2)}{(-u^2+v^2)}\log\left|\frac{u^2-1}{v^2-1}\right|+2\right] \ .
\end{align}

\begin{figure}[H]
    \centering
    \begin{subfigure}{0.49\textwidth}
        \includegraphics[width=\linewidth]{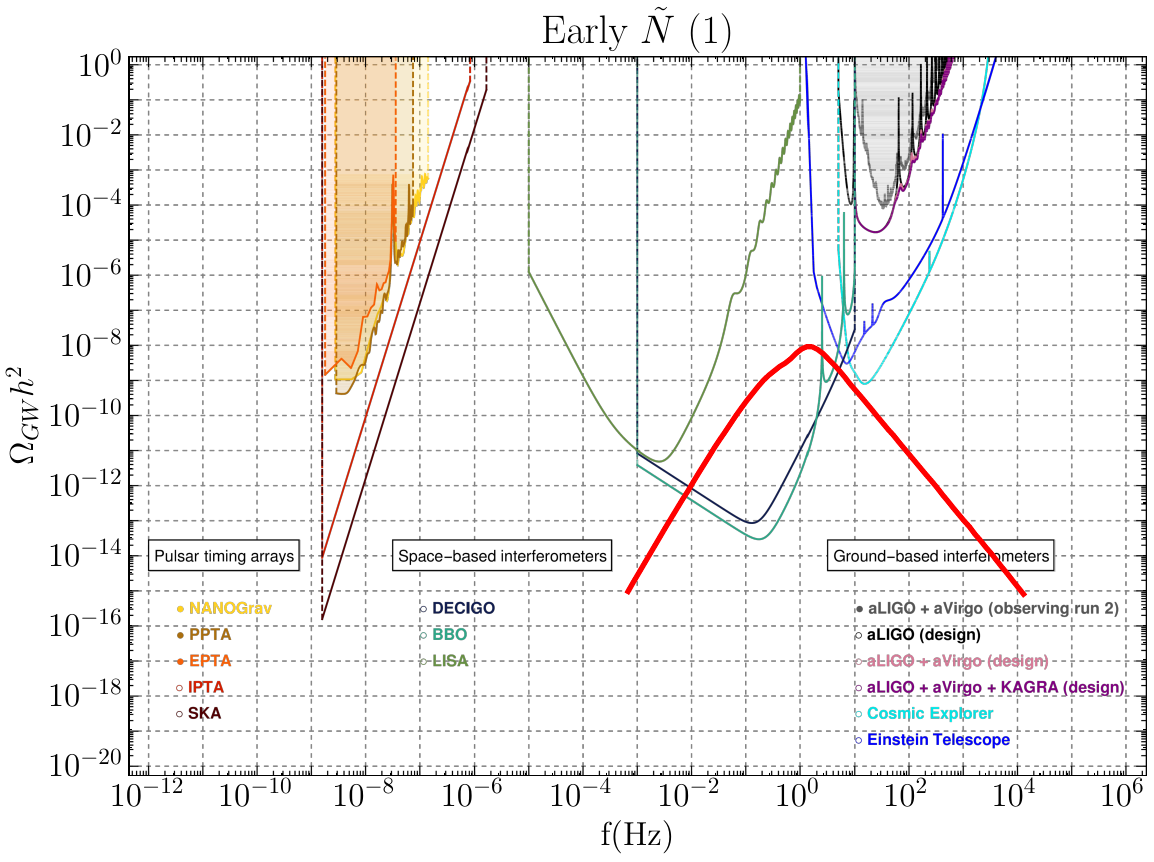}
    \end{subfigure}
    \begin{subfigure}{0.49\textwidth}
        \includegraphics[width=\linewidth]{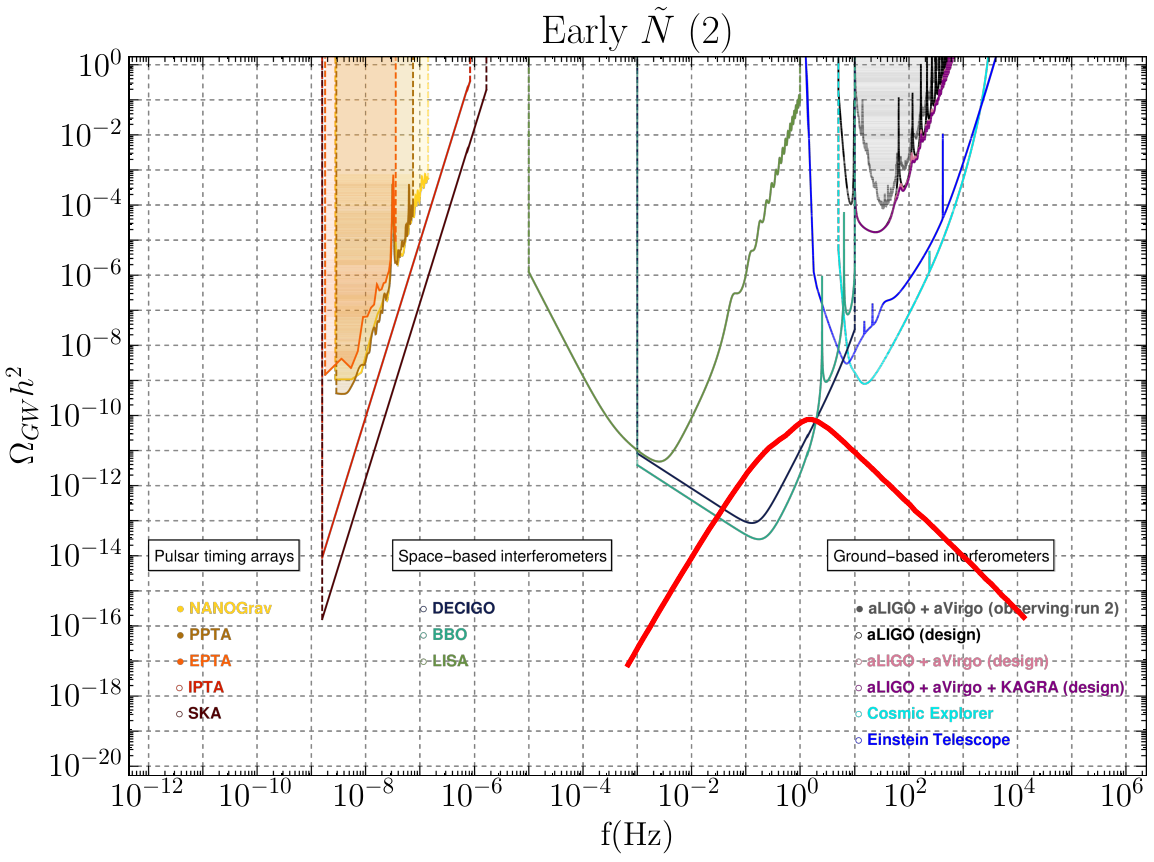}
    \end{subfigure}
    \vspace{\baselineskip}
    \begin{subfigure}{0.49\textwidth}
        \includegraphics[width=\linewidth]{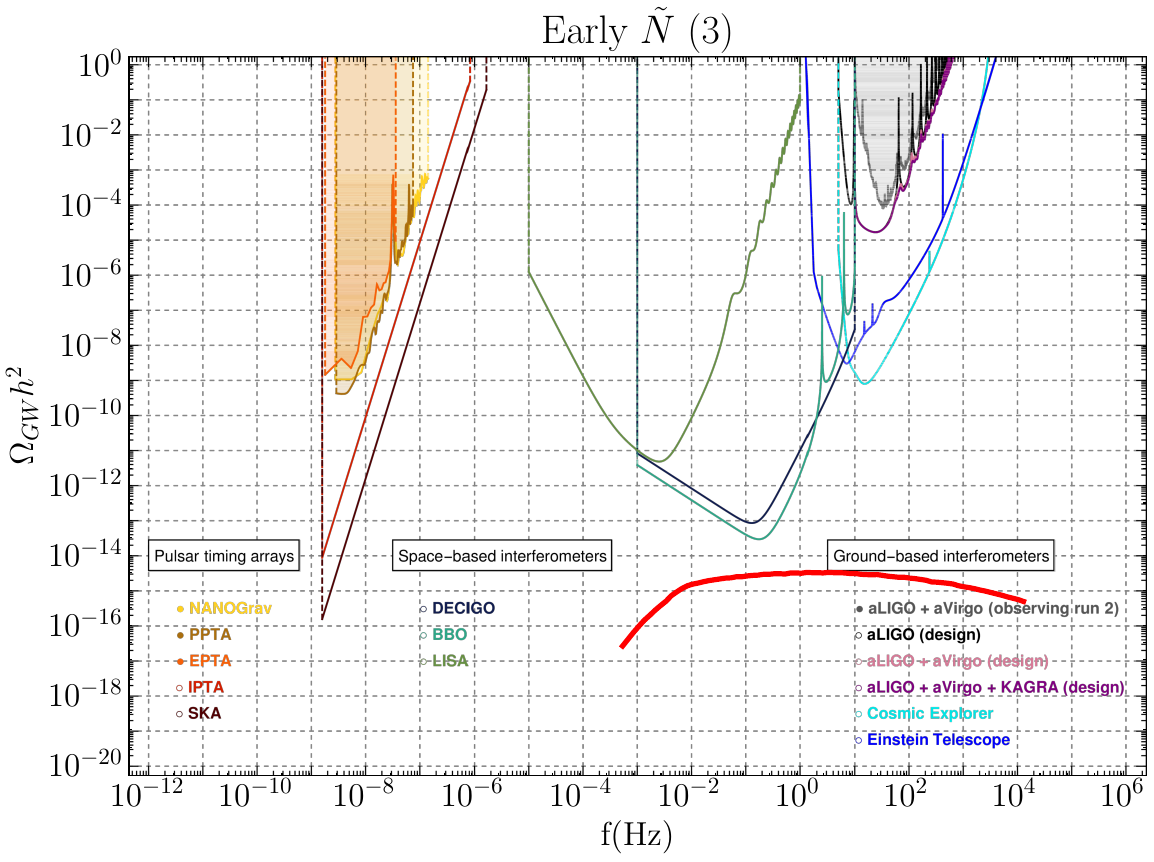}
    \end{subfigure}
    \begin{subfigure}{0.49\textwidth}
        \includegraphics[width=\linewidth]{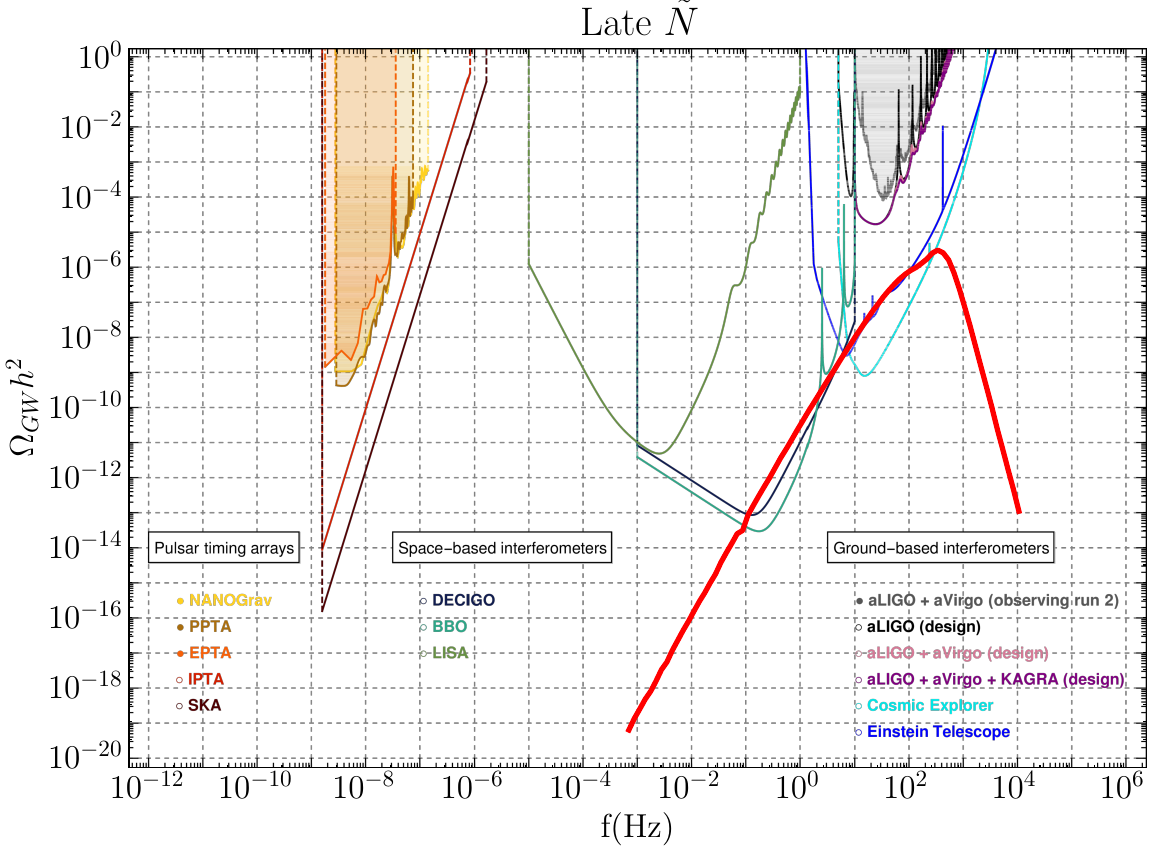}
    \end{subfigure}
    \caption{In red: gravitational wave spectrum for the points in~\cref{tab:tweaked_good_points}. The sensitivity curves are from~\cite{Schmitz:2020syl}.}
    \label{fig:omegagw_vs_f_multiple_points}
\end{figure}

For each point in~\cref{tab:tweaked_good_points}, we compute~\cref{eq:omegagw} for $k$ in the range $[k(30),k_{max}]$ where $k_{max} = \min(k(\Delta N), 10^{19} \ \text{Mpc}^{-1})$, and this way we cover the range where the power spectrum is amplified, c.f.~\cref{fig:PR_vs_PR_SR_multiple_points}. The present day gravitational spectrum, as a function of the wavelength, is presented in~\cref{fig:omegagw_vs_f_multiple_points} for the points presented in~\cref{tab:tweaked_good_points}.
We observe that, for earlier peaks with enough power spectrum enhancement, the produced gravitational waves are within the detectable range of space-based interferometers, more concretely BBO and DECIGO. On the other hand, for later peaks, the gravitational wave signal is also in the range of ground-based interferometers, such as the Cosmic Explorer and the Einstein Telescope, whilst also being detectable by the mentioned space-based interferometers. Finally, for points with a wide power spectrum enhancement, identified as those with a long plateau appearing earlier during inflation, the gravitational wave signal is expected to evade any possible detection on current and near-future interferometers.

\subsection{The Effect of the Gravitino Mass on the Gravitational Waves Spectrum\label{sec:gw_and_m32}}

In~\cref{sec:scan} we studied how the different parameters of the potential impact the Planck constrained observables and the gravitino mass. It was shown that the gravitino mass is largely insensitive to the parameters arising from the \kahler\ potential. On the other hand, these are the parameters that govern the shape and the size of the kink which ultimately drive the enhancement of the power spectrum that source the production of the GW. Whilst the discussion so far has concerned itself on how the presence of the kink might affect the value of the gravitino mass, we have not yet discussed the converse statement: how the presence of a non-vanishing gravitino mass impacts the phenomenology of the GW.

To explore this, we select one of the points from~\cref{tab:tweaked_good_points} studied above. More precisely, we choose the second point with early power spectrum enhancement, which we have labelled as Early $\tilde N (2)$. This point has the lowest value of $d$, at $d=9.25 \times 10^{-7}$, and consequently the lowest gravitino mass from the set, c.f.~\cref{tab:tweaked_points_observables}. We will now change the value of $d$ to take the following values\footnote{We cannot take $d$ to be much greater than $10^{-6}$ for this point as it leads to unsatisfactory inflation.}
\begin{align}
    d \in \{0,\ 1.0\times 10^{-7},\ 2.5.0\times 10^{-7},\ 5.0\times 10^{-7},\ 7.5\times 10^{-7},\ 1.0\times 10^{-6}\} \ ,
    \label{eq:d_range}
\end{align}
and for each case we compute the resulting power spectrum and the GW energy density spectrum.
\begin{figure}[H]
    \centering
    \includegraphics[width=1.0\textwidth]{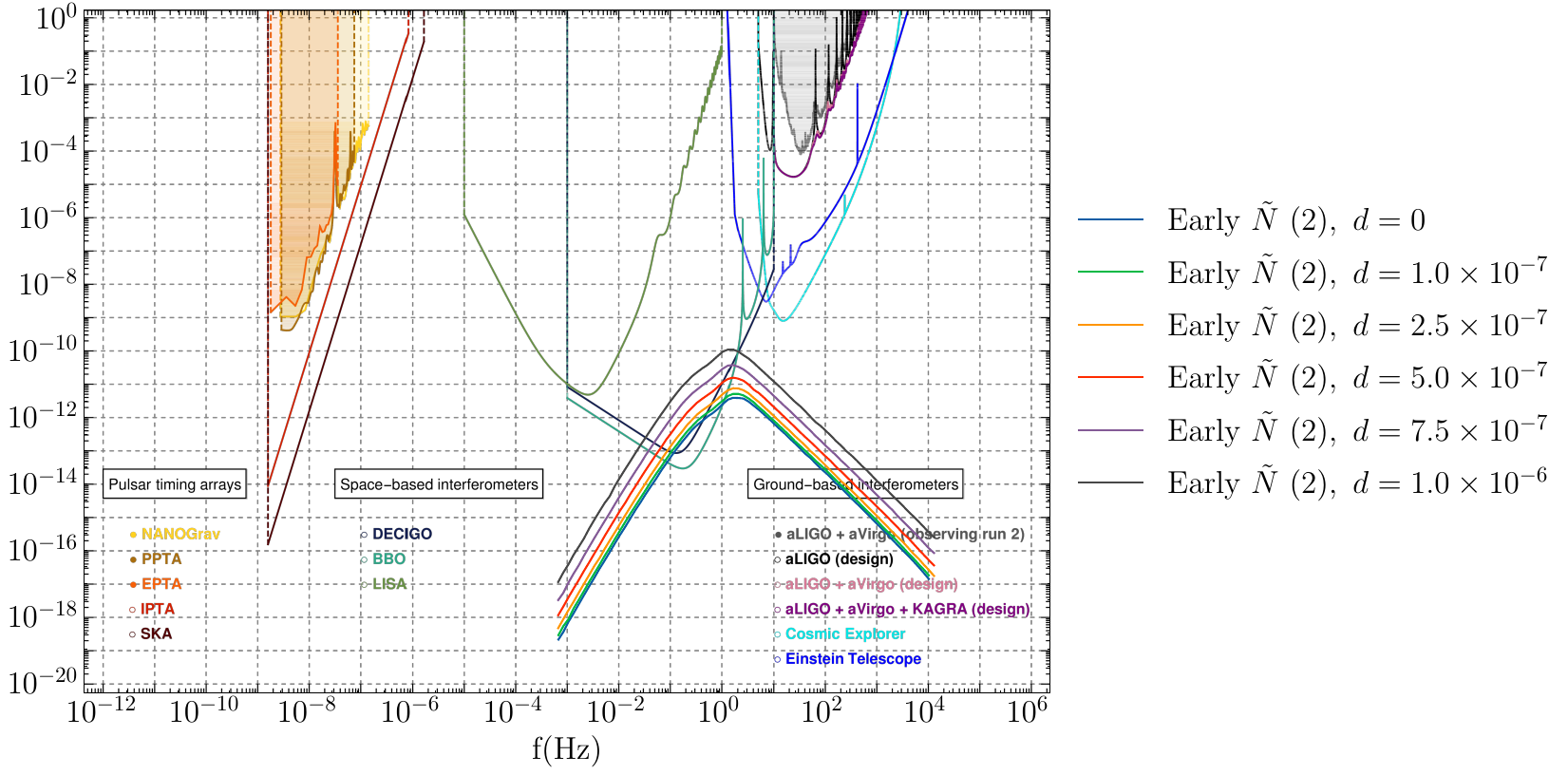}
    \caption{Gravitational wave spectrum for the second point from~\cref{tab:tweaked_good_points}, with $d$ taking values from~\cref{eq:d_range}. The sensitivity curves are from~\cite{Schmitz:2020syl}.}
    \label{fig:omegagw_vs_f_Early 2_varying_d}
\end{figure}
\begin{table}[H]
    \centering
    \begin{tabular}{ccccc}
        $d$                 & $n_s$    & $r$       & $m_{3/2}$ (GeV)  & $\Delta N$ \\
        \hline
        \hline
        $0$                 & $0.9566$ & $0.00902$ & $0.0$            & $52.4$     \\
        $1.0\times 10^{-7}$ & $0.9567$ & $0.00906$ & $8.9\times10^1$  & $52.5$     \\
        $2.5\times 10^{-7}$ & $0.9568$ & $0.00912$ & $5.6\times10^2$  & $52.7$     \\
        $5.0\times 10^{-7}$ & $0.9571$ & $0.00921$ & $2.2\times 10^3$ & $53.1$     \\
        $7.5\times 10^{-7}$ & $0.9574$ & $0.00930$ & $5.1\times 10^3$ & $53.6$     \\
        $1.0\times 10^{-6}$ & $0.9576$ & $0.00940$ & $9.0\times 10^3$ & $54.2$
    \end{tabular}
    \caption{Impact on the inflation observables, gravitino mass, and number of e-folds by changing the parameter the value of the parameter $d$ for the second point in~\cref{tab:tweaked_points_observables} over~\cref{eq:d_range}.}
    \label{tab:observables_varying_d}
\end{table}

For each value of $d$ in the range~\cref{eq:d_range}, we show the resulting GW energy density spectrum
in~\cref{fig:omegagw_vs_f_Early 2_varying_d}, and  the inflationary observables, the gravitino mass, and the number of e-folds in~\cref{tab:observables_varying_d}. The results show a somehow unexpected phenomenon: as $d$ increases the GW energy density increases. Since $d$ drives the gravitino mass, one can then conclude that for a given combination of parameters, a higher gravitino mass will be associated with a higher GW energy density. Notice that GW would still be detectable in the limit of a massless gravitino, i.e. for $d=0$. While the existence of the kink, and consequently of observable GW, is governed by the \kahler\ potential parameters $a$, $b$, and $c$, the total energy density of the GW spectrum can be boosted by the Polonyi term, governed by $d$, which also drives the value of the gravitino mass. From~\cref{tab:observables_varying_d} we can see that increasing $d$ has a similar effect as decreasing $b$ through tuning: the inflaton \emph{spends more time} traversing the plateau, leading to a slightly higher number of e-folds, but in a far less dramatic way than when tuning $b$. Also from the same table, we see how the Planck constrained observables are minimally altered, as expected from the discussion above and from our previous work. Therefore, while a GW signal is still dependent on fine tuning the parameter $b$, as it was observed and discussed in~\cite{Stamou:2021qdk}, our model provides a new source of GW energy density enhancement via the Polonyi term with reduced fine tuning.

\section{Conclusions\label{sec:conclsions}}

In this work we studied the production of GW in No-Scale SUGRA inflation. In~\cref{sec:noscale} we presented the main No-Scale SUGRA inflation framework. We revisited our previous model, where the Wess-Zumino superpotential is extended to include a Polonyi term, and further extended it to include a non-minimal \kahler\ superpotential, which describes a deformation of the original space described by a non-compact $SU(2,1)/SU(2)\times U(1)$ coset. This deformation leads to a kink feature in the effective scalar potential of the inflaton, which, for certain values of the parameters, exhibits an inflexion point where the inflaton slows down. This slowing down enhances the power spectrum, which in turn can source the production of GW at later stages of inflation.

After discussing how the shape of the inflaton potential changes for different values of the parameters in~\cref{sec:V}, we proceeded to study the resulting inflationary dynamics across the parameter space by performing a thorough scan in~\cref{sec:scan}. We observed that there are regions of the parameter space, for $a\ll 0$ (with large $|a|$) and small $b$, where the inflaton can get stuck at the inflexion point where the kink feature becomes effectively a wall. This reinforces the bounds on the \kahler\ potential parameters $a\gtrsim-1.5$ and $b\gtrsim15$, which were previously identified to ensure that the potential is positive semi-definite with a global minimum at $V=0$. Focusing on the points that allow for the inflaton to roll down to the global minimum, we obtained an upper bound on the parameter $d$ that sets scale of the Polonyi term for the points that pass the Planck satellite constraints. This parameter is responsible for SUSY breaking at the end of inflation, producing a gravitino mass that is bounded from above to be at most $\mathcal{O}(1000)$ TeV for the region of the parameter space that produces the desirable inflationary dynamics, in agreement with our previous work. As in our previous work, this means that the SUSY scale cannot be exceed the PeV scale, and therefore that SUSY should eventually be discovered.

Having identified the parameter space points that produce good inflationary dynamics and observables, we then turned to the study of the power spectrum. We first performed an extensive analysis of the power spectrum in~\cref{sec:exploratory_PR} using the slow-roll approximation. We found that in general the points that exhibit a power spectrum peak around the time that the inflaton traverses the inflexion point fall short in enhancing the power spectrum for the production of observable GW. We then investigated whether this generic expectation of unobservable GW is a robust conclusion, or whether it is indeed possible for the model to produce an observable GW signal is certain corners of parameter space.
Since it is known from the literature that the parameter $b$ can be tuned to augment the power spectrum peak around the inflexion point by slowing down the inflaton, we then changed our strategy and focused on points that pass the Planck satellite constraints, but end inflation earlier than desired, $\Delta N < 50$. From this set of points, we further identified three promising points as those for which the power spectrum is narrowly peaked, corresponding to $a \lesssim - 0.9$, i.e., $|a| \simeq 1$. We were able to tune the parameter $b$ of these four points, obtaining a pronounced power spectrum peak enhancement, while producing $50 \le\Delta N \le 60$. These three points also produce a relatively earlier peak of the power spectrum. To complement these points, we also identified a fourth point with a later power spectrum.

We then proceeded to compute the full numerical solution for the power spectrum and the resulting GW energy density spectrum for these four points in~\cref{sec:PR_and_GW}. In~\cref{sec:PR} we confirmed the result that a crucial characteristic to have an enhancement of the power spectrum is for its peak to be narrow, with a sharp increase. For the three points where the power spectrum exhibits this characteristic, the full numerical solution produces a peak at least an order of magnitude higher than the one suggested by the slow-roll approximation. We then proceeded to study the resulting GW spectrum in~\cref{sec:GW}, where we found that for the points with sufficiently enhanced power spectrum peak, that a GW signal can be detectable by near future interferometers. More precisely, for the points with an earlier peak, $\tilde N\lesssim 40$ the spectrum falls within the detection range of space-based interferometers BBO and DECIGO, while for the point with a later peak, $\tilde N \simeq 45$, it would also be possible to detect the signal in ground-based interferometers such as the Cosmic Explorer and the Einstein Telescope.

In~\cref{sec:gw_and_m32} we performed an analysis of how the presence of the Polonyi term would affect the  GW energy density spectrum. Choosing one of the four points identified before, we varied the value of the parameter $d$ governing the magnitude of the Polonyi term, and found that the GW energy density is augmented by the presence of a non-vanishing Polonyi term. While our result does not cure the high level of tuning associated with the parameter $b$, it does provide a further mechanism to increase the GW signal in near future interferometers. Notwithstanding that the main parameter governing the power spectrum peak and the resulting GW spectrum is a highly tuned $b$, our result still suggests an interplay between the detection of GW and the discovery of SUSY at current and near future colliders.

Although our analysis focuses on No-Scale SUGRA inflationary models with an inflexion point arising from a non-minimal \kahler\ potential, it would be an interesting avenue of research to assess whether more general inflationary potentials with similar features lead to equivalent findings. Furthermore, as an enhanced power spectrum peak is also known to lead to the production of PBH, which can contribute to both the NANOGrav observed GW cosmic background, but also to the Dark Matter relic density, it would be interesting to study whether there is an interplay between the gravitino mass and the Dark Matter phenomenology. We leave these open questions for future work.

\section*{Acknowledgements}

We are immensely grateful to Dr. Ioanna Stamou for her patience in helping us to validate our methodology and to reproduce the results in the literature.
SFK acknowledges the STFC Consolidated Grant ST/L000296/1 and the European Union's Horizon 2020 Research and Innovation programme under Marie Sklodowska-Curie grant agreement HIDDeN European ITN project (H2020-MSCA-ITN-2019//860881-HIDDeN).

\bibliography{paper}{}

\begin{thebibliography}{10}

\bibitem{LIGOScientific:2016aoc}
B.~P. Abbott et~al.
\newblock {Observation of Gravitational Waves from a Binary Black Hole Merger}.
\newblock {\em Phys. Rev. Lett.}, 116(6):061102, 2016.

\bibitem{LIGOScientific:2016sjg}
B.~P. Abbott et~al.
\newblock {GW151226: Observation of Gravitational Waves from a 22-Solar-Mass Binary Black Hole Coalescence}.
\newblock {\em Phys. Rev. Lett.}, 116(24):241103, 2016.

\bibitem{LIGOScientific:2017ycc}
B.~P. Abbott et~al.
\newblock {GW170814: A Three-Detector Observation of Gravitational Waves from a Binary Black Hole Coalescence}.
\newblock {\em Phys. Rev. Lett.}, 119(14):141101, 2017.

\bibitem{NANOGRAV:2018hou}
Z.~Arzoumanian et~al.
\newblock {The NANOGrav 11-year Data Set: Pulsar-timing Constraints On The Stochastic Gravitational-wave Background}.
\newblock {\em Astrophys. J.}, 859(1):47, 2018.

\bibitem{NANOGrav:2020bcs}
Zaven Arzoumanian et~al.
\newblock {The NANOGrav 12.5 yr Data Set: Search for an Isotropic Stochastic Gravitational-wave Background}.
\newblock {\em Astrophys. J. Lett.}, 905(2):L34, 2020.

\bibitem{LISA:2017pwj}
Pau Amaro-Seoane et~al.
\newblock {Laser Interferometer Space Antenna}.
\newblock 2 2017.

\bibitem{phinney2004big}
Sterl Phinney, Peter Bender, R~Buchman, Robert Byer, Neil Cornish, Peter Fritschel, and S~Vitale.
\newblock The big bang observer: Direct detection of gravitational waves from the birth of the universe to the present.
\newblock {\em NASA mission concept study}, 2004.

\bibitem{Sato:2017dkf}
Shuichi Sato et~al.
\newblock {The status of DECIGO}.
\newblock {\em J. Phys. Conf. Ser.}, 840(1):012010, 2017.

\bibitem{Reitze:2019iox}
David Reitze et~al.
\newblock {Cosmic Explorer: The U.S. Contribution to Gravitational-Wave Astronomy beyond LIGO}.
\newblock {\em Bull. Am. Astron. Soc.}, 51(7):035, 2019.

\bibitem{Maggiore:2019uih}
Michele Maggiore et~al.
\newblock {Science Case for the Einstein Telescope}.
\newblock {\em JCAP}, 03:050, 2020.

\bibitem{Acquaviva:2002ud}
Viviana Acquaviva, Nicola Bartolo, Sabino Matarrese, and Antonio Riotto.
\newblock {Second order cosmological perturbations from inflation}.
\newblock {\em Nucl. Phys. B}, 667:119--148, 2003.

\bibitem{Ananda:2006af}
Kishore~N. Ananda, Chris Clarkson, and David Wands.
\newblock {The Cosmological gravitational wave background from primordial density perturbations}.
\newblock {\em Phys. Rev. D}, 75:123518, 2007.

\bibitem{Baumann:2007zm}
Daniel Baumann, Paul~J. Steinhardt, Keitaro Takahashi, and Kiyotomo Ichiki.
\newblock {Gravitational Wave Spectrum Induced by Primordial Scalar Perturbations}.
\newblock {\em Phys. Rev. D}, 76:084019, 2007.

\bibitem{Espinosa:2018eve}
Jos\'e~Ram\'on Espinosa, Davide Racco, and Antonio Riotto.
\newblock {A Cosmological Signature of the SM Higgs Instability: Gravitational Waves}.
\newblock {\em JCAP}, 09:012, 2018.

\bibitem{Nanopoulos:2020nnh}
Dimitri~V. Nanopoulos, Vassilis~C. Spanos, and Ioanna~D. Stamou.
\newblock {Primordial Black Holes from No-Scale Supergravity}.
\newblock {\em Phys. Rev. D}, 102(8):083536, 2020.

\bibitem{Stamou:2021qdk}
Ioanna~D. Stamou.
\newblock {Mechanisms of producing primordial black holes by breaking the $SU(2, 1)/SU(2)\times U(1)$ symmetry}.
\newblock {\em Phys. Rev. D}, 103(8):083512, 2021.

\bibitem{Spanos:2022euu}
Vassilis~C. Spanos and Ioanna~D. Stamou.
\newblock {Gravitational waves from no-scale supergravity}.
\newblock {\em Eur. Phys. J. C}, 83(1):4, 2023.

\bibitem{Guth:1980zm}
Alan~H. Guth.
\newblock {The Inflationary Universe: A Possible Solution to the Horizon and Flatness Problems}.
\newblock {\em Phys. Rev. D}, 23:347--356, 1981.

\bibitem{Linde:1981mu}
Andrei~D. Linde.
\newblock {A New Inflationary Universe Scenario: A Possible Solution of the Horizon, Flatness, Homogeneity, Isotropy and Primordial Monopole Problems}.
\newblock {\em Phys. Lett. B}, 108:389--393, 1982.

\bibitem{Mukhanov:1981xt}
Viatcheslav~F. Mukhanov and G.~V. Chibisov.
\newblock {Quantum Fluctuations and a Nonsingular Universe}.
\newblock {\em JETP Lett.}, 33:532--535, 1981.

\bibitem{Albrecht:1982wi}
Andreas Albrecht and Paul~J. Steinhardt.
\newblock {Cosmology for Grand Unified Theories with Radiatively Induced Symmetry Breaking}.
\newblock {\em Phys. Rev. Lett.}, 48:1220--1223, 1982.

\bibitem{Linde:1983gd}
Andrei~D. Linde.
\newblock {Chaotic Inflation}.
\newblock {\em Phys. Lett. B}, 129:177--181, 1983.

\bibitem{Linde:2007fr}
Andrei~D. Linde.
\newblock {Inflationary Cosmology}.
\newblock {\em Lect. Notes Phys.}, 738:1--54, 2008.

\bibitem{Linde:1990flp}
Andrei~D. Linde.
\newblock {\em {Particle physics and inflationary cosmology}}, volume~5.
\newblock 1990.

\bibitem{Lyth:1998xn}
David~H. Lyth and Antonio Riotto.
\newblock {Particle physics models of inflation and the cosmological density perturbation}.
\newblock {\em Phys. Rept.}, 314:1--146, 1999.

\bibitem{Planck:2018jri}
Y.~Akrami et~al.
\newblock {Planck 2018 results. X. Constraints on inflation}.
\newblock {\em Astron. Astrophys.}, 641:A10, 2020.

\bibitem{Ringeval:2007am}
Christophe Ringeval.
\newblock {The exact numerical treatment of inflationary models}.
\newblock {\em Lect. Notes Phys.}, 738:243--273, 2008.

\bibitem{Starobinsky:1980te}
Alexei~A. Starobinsky.
\newblock {A New Type of Isotropic Cosmological Models Without Singularity}.
\newblock {\em Phys. Lett. B}, 91:99--102, 1980.

\bibitem{Starobinsky:1983zz}
A.~A. Starobinsky.
\newblock {The Perturbation Spectrum Evolving from a Nonsingular Initially De-Sitter Cosmology and the Microwave Background Anisotropy}.
\newblock {\em Sov. Astron. Lett.}, 9:302, 1983.

\bibitem{Bezrukov:2009db}
F.~Bezrukov and M.~Shaposhnikov.
\newblock {Standard Model Higgs boson mass from inflation: Two loop analysis}.
\newblock {\em JHEP}, 07:089, 2009.

\bibitem{Linde:2011nh}
Andrei Linde, Mahdiyar Noorbala, and Alexander Westphal.
\newblock {Observational consequences of chaotic inflation with nonminimal coupling to gravity}.
\newblock {\em JCAP}, 03:013, 2011.

\bibitem{Ferrara:2010in}
Sergio Ferrara, Renata Kallosh, Andrei Linde, Alessio Marrani, and Antoine Van~Proeyen.
\newblock {Superconformal Symmetry, NMSSM, and Inflation}.
\newblock {\em Phys. Rev. D}, 83:025008, 2011.

\bibitem{Copeland:1994vg}
Edmund~J. Copeland, Andrew~R. Liddle, David~H. Lyth, Ewan~D. Stewart, and David Wands.
\newblock {False vacuum inflation with Einstein gravity}.
\newblock {\em Phys. Rev. D}, 49:6410--6433, 1994.

\bibitem{Dvali:1994ms}
G.~R. Dvali, Q.~Shafi, and Robert~K. Schaefer.
\newblock {Large scale structure and supersymmetric inflation without fine tuning}.
\newblock {\em Phys. Rev. Lett.}, 73:1886--1889, 1994.

\bibitem{Ellis:1982ed}
John~R. Ellis, Dimitri~V. Nanopoulos, Keith~A. Olive, and K.~Tamvakis.
\newblock {Cosmological Inflation Cries Out for Supersymmetry}.
\newblock {\em Phys. Lett. B}, 118:335, 1982.

\bibitem{Ellis:1982dg}
John~R. Ellis, Dimitri~V. Nanopoulos, Keith~A. Olive, and K.~Tamvakis.
\newblock {Fluctuations in a Supersymmetric Inflationary Universe}.
\newblock {\em Phys. Lett. B}, 120:331--334, 1983.

\bibitem{Ellis:1982ws}
John~R. Ellis, Dimitri~V. Nanopoulos, Keith~A. Olive, and K.~Tamvakis.
\newblock {PRIMORDIAL SUPERSYMMETRIC INFLATION}.
\newblock {\em Nucl. Phys. B}, 221:524--548, 1983.

\bibitem{Ellis:2020lnc}
John Ellis, Marcos A.~G. Garcia, Natsumi Nagata, Nanopoulos~Dimitri V., Keith~A. Olive, and Sarunas Verner.
\newblock {Building models of inflation in no-scale supergravity}.
\newblock {\em Int. J. Mod. Phys. D}, 29(16):2030011, 2020.

\bibitem{Ellis2013-hj}
John Ellis, Dimitri~V Nanopoulos, and Keith~A Olive.
\newblock No-scale supergravity realization of the starobinsky model of inflation.
\newblock {\em Phys. Rev. Lett.}, 111(11):111301, September 2013.

\bibitem{Ellis:2013nxa}
John Ellis, Dimitri~V. Nanopoulos, and Keith~A. Olive.
\newblock {Starobinsky-like Inflationary Models as Avatars of No-Scale Supergravity}.
\newblock {\em JCAP}, 10:009, 2013.

\bibitem{Romao2017-pj}
Miguel~Crispim Romao and Stephen~F King.
\newblock Starobinsky-like inflation in no-scale supergravity {Wess-Zumino} model with polonyi term.
\newblock {\em JHEP}, 07:033, March 2017.

\bibitem{King:2019omb}
Stephen~F. King and Elena Perdomo.
\newblock {Starobinsky-like inflation and soft-SUSY breaking}.
\newblock {\em JHEP}, 05:211, 2019.

\bibitem{Forster:2021vyz}
Adam~K. Forster and Stephen~F. King.
\newblock {Muon g-2, dark matter and the Higgs mass in no-scale supergravity}.
\newblock {\em Nucl. Phys. B}, 976:115700, 2022.

\bibitem{Martin1997-bs}
Stephen~P Martin.
\newblock A supersymmetry primer.
\newblock September 1997.

\bibitem{Brignole1997-vd}
A~Brignole, L~E Iba{\~n}ez, and C~Mu{\~n}oz.
\newblock Soft supersymmetry-breaking terms from supergravity and superstring models.
\newblock July 1997.

\bibitem{Spanos:2021hpk}
Vassilis~C. Spanos and Ioanna~D. Stamou.
\newblock {Gravitational waves and primordial black holes from supersymmetric hybrid inflation}.
\newblock {\em Phys. Rev. D}, 104(12):123537, 2021.

\bibitem{Schmitz:2020syl}
Kai Schmitz.
\newblock {New Sensitivity Curves for Gravitational-Wave Signals from Cosmological Phase Transitions}.
\newblock {\em JHEP}, 01:097, 2021.

\end{thebibliography}
\bibliographystyle{unsrt}
\end{document}